\documentclass[11pt,a4paper]{article}

\usepackage{jcappub}
\usepackage{aas_macros}

\newcommand{\be}{\begin{equation}}
\newcommand{\ee}{\end{equation}}           
\newcommand{\fnl}{f_{\rm{NL}}}

\def\beq{\begin{equation}}
\def\eeq{\end{equation}}
\def\baq{\begin{eqnarray}}
\def\eaq{\end{eqnarray}}

\def\ka{\kappa}
\def\fnl{f_{\rm NL}}

\def\fnl{f_{\rm NL}}

\def\bea{\begin{eqnarray}}
\def\eea{\end{eqnarray}}
\def\be{\begin{equation}}
\def\ee{\end{equation}}

\begin{document}

\setcounter{page}{0}

\title{Mixed non-Gaussianity in multiple-DBI inflation}

\author{Jon Emery,}
\author{Gianmassimo Tasinato}
\author{and David Wands}

\affiliation{Institute of Cosmology \& Gravitation, University of Portsmouth,}
\affiliation{Dennis Sciama Building, Portsmouth, PO1 3FX, United Kingdom}

\emailAdd{jon.emery@port.ac.uk}
\emailAdd{gianmassimo.tasinato@port.ac.uk}
\emailAdd{david.wands@port.ac.uk}

\abstract{We study a model of multiple-field DBI inflation leading to mixed form
of primordial non-Gaussianity, including equilateral and local
bispectrum shapes. We present a general formalism based on the
Hamilton-Jacobi approach, allowing us to go beyond slow-roll,
combining the three-point function for the fields at Hubble-exit with
the non-linear evolution of super-Hubble scales. We are able to obtain
analytic results by taking a separable Ansatz for the Hubble rate. We
find general expressions for both the equilateral and local type
non-Gaussianity parameter $\fnl$. The equilateral non-Gaussianity
includes the usual enhancement for small sound speeds, but multiplied
by an analytic factor which can lead to a suppression. We illustrate
our results with two scenarios. In the first model, previously found
to have detectable local non-Gaussianity, we find that the equilateral
signal is not sufficiently suppressed to evade current observational
bounds. In our second scenario we construct a model which exhibits
both a detectable equilateral $\fnl$ and a negative local $\fnl$.}


\notoc

\maketitle

\newpage

\setcounter{page}{1}

\section{Introduction\label{sec:introduction}}

Inflation is widely believed to be responsible for the specific set of initial conditions on which the hot big bang relies. Whilst a compelling mechanism, a consistent model proves elusive however (see \cite{Lyth99,Lyth09,Mazumdar10} for reviews), which is in part due to the limited information available in the two-point statistics of the primordial density perturbations. Potential non-Gaussian signatures have therefore become an increasingly popular observable with which to discern between otherwise degenerate models \cite{Komatsu09}, particularly since impending observations are set to improve by at least an order of magnitude \cite{Planck06}. It is therefore important to try to understand the correspondence between inflationary dynamics and the different forms of non-Gaussianity (see \cite{Bartolo04,Chen10} for recent reviews).

Non-Gaussianity can be produced by inflation in a number of distinct ways. For example, by converting between entropy and adiabatic modes during\footnote{Non-linearities can also develop through the conversion between entropy and adiabatic modes after inflation. For example, the curvaton mechanism \cite{Moroi01,Lyth02,Lyth03} and modulated reheating \cite{Kofman03,Dvali04}. See
\cite{Burgess10,Cicoli12} for recent realisations of such scenarios in string theory.} multiple-field inflation \cite{Gordon01,Nibbelink02,Rigopoulos04}, the curvature perturbation $\zeta$ can evolve on super-horizon scales \cite{Wands00,Lyth05}. Such non-linearities can in principle produce \emph{local-type} non-Gaussianity \cite{Bernardeau02,Vernizzi06,Rigopoulos06,Rigopoulos07,Yokoyama07,Battefeld07,Yokoyama08,Byrnes08,Sasaki08,Byrnes09,Battefeld09,Wang10,Tzavara11,Elliston11a,Elliston11b,Meyers11-a,Meyers11-b,Peterson11a,Watanabe11,Choi12,Mazumdar12,Frazer11a,Battefeld12}, where the effect is associated with a turn in the trajectory and is often enhanced by violations of slow roll. Alternatively, single field models with non-standard kinetic terms provide an alternative source of non-Gaussianity (see \cite{Chen07,Koyama10} and references therein). Often motivated by string theory, the models we are concerned with have a characteristic sound speed $c_{s}$, where $c_{s}=1$ in the canonical case.\footnote{Although generally referred to as the sound speed this is technically the phase speed of fluctuations. See \cite{Christopherson09} for further clarification of this point.} As a result, \emph{equilateral} non-Gaussianity can be produced by the interactions of quantum fields on sub-horizon scales. This is the case in Dirac-Born-Infeld (DBI) inflation \cite{Silverstein04,Alishahiha04}, in which a probe D-brane moving along the radial direction of a warped throat drives inflation.  

More generally however, it is expected that both equilateral and local contributions will be relevant in models characterised by non-standard kinetic terms and multiple-field dynamics \cite{Langlois08,Langlois08b,Arroja08,RenauxPetel09a,Gao09,Mizuno09,Langlois09a,Gao09a,Cai09,Cai09-1,Pi11,Emery12,Kidani12}.  For example, in previous work \cite{Emery12} we studied a multiple-DBI model, akin to that of \cite{Cai09,Cai09-1,Pi11}, as a concrete example of multi-component inflation with non-standard kinetic terms. Using the $\delta N$ formalism, we tracked the super-horizon evolution of perturbations using the field fluctuations at horizon exit and the subsequent background trajectory. With the adoption of a sum separable Hubble parameter, as in \cite{Byrnes09}, we were able to treat the two-field case both analytically and beyond slow variation to calculate the local contribution. Moreover, by considering inflation in the tip regions of two warped throats, we illustrated that rapidly varying sound speeds can produce large local type non-Gaussianity during a turn in the trajectory.

Our previous work did not include the equilateral contribution produced on sub-horizon scales however, which is what we intend to address in this paper. Whilst this contribution is dominant in the single field case, the introduction of multiple  fields can alter this conclusion through the conversion of entropy and adiabatic modes (see, for example, \cite{Langlois08,Langlois08b,Arroja08}). In this paper we again consider the multiple-DBI model as in \cite{Emery12} and compute the full third order action using the Arnowitt-Deser-Misner (ADM) formalism \cite{Arnowitt62}. After considering the leading contributions in slow variation and small sound speeds, we calculate the three point function for the field fluctuations at horizon exit using the path integral approach \cite{Seery05b}. Thereafter, we implement the $\delta N$ formalism and assume a separable Hubble parameter as in our previous work to calculate, fully analytically, the combined local and equilateral contributions to the bispectrum of the curvature perturbation, giving one of the few explicit examples of models characterised by both contributions (see \cite{RenauxPetel09,Kidani12} for alternatives). Finally, as a first step towards assessing the viability of such a signal, we apply our results to two specific cases.  

The outline of the paper is as follows. We begin in section~\ref{sec:model} by briefly reviewing the multiple-DBI model and introducing and reformulating the relevant quantities using the $\delta N$ formalism. Thereafter, we use the path integral method to calculate the three point function of field fluctuations at horizon crossing in section~\ref{sec:three_point}, having first derived the third order action for this scenario. We then use the $\delta N$ formalism to present the corresponding equilateral non-linearity parameter. In section~\ref{sec:full_three_point} we assume a separable Hubble parameter to combine this result with the local contribution found in our previous work, giving the total three point function for the curvature perturbation. We briefly assess the feasibility of such a signal by studying two specific examples in section~\ref{sec:models}. Finally, we conclude in section~\ref{sec:conclusions}.

Throughout this paper we use the $(-,+,+,+)$ metric signature and set $M_{\textrm{P}}=c=1$, where $M_{\textrm{P}}=\frac{1}{\sqrt{8\pi G}}$ is the reduced Planck mass. Capital latin indices label scalar fields and any summation is explicit. Greek indices label space-time co-ordinates whilst lower case latin indices label spatial co-ordinates only, where the Einstein summation convention is adopted. Finally, commas denote partial derivatives and over-dots represent derivatives with respect to cosmic time. 

\section{Multiple-DBI inflation and non-Gaussianity\label{sec:model}}

We begin this section by briefly reviewing the multiple-DBI model and the relevant observational quantities, paying particular attention to non-Gaussianity. Thereafter, we use the $\delta N$ formalism to re-write these expressions  
and investigate their evolution on super-horizon scales. This section is intended to be relatively brief, since full details can be found in \cite{Emery12}.

\subsection{Background evolution in multiple-DBI\label{sec:model_homo}}

Multiple-DBI inflation is encompassed by the following action

      \be S=\frac{1}{2} \int d^{4}x\sqrt{-g}\bigl[R+2\sum_{I}P_{I}-2V\bigr],\label{eq:action-general}\ee

\noindent where $P_{I}$ is a function of the single scalar field $\phi_{I}$ and kinetic function $X_{I}=-\frac{1}{2}g^{\mu\nu}\phi_{I,\mu}\phi_{I,\nu}$, whilst the potential $V$ is a function of the set of scalar fields $\phi=\{ \phi_{1},\phi_{2},...,\phi_{n}\}$. We model inflation through $n$ probe D3 branes descending $n$ distinct warped throats glued to a compact Calabi-Yau manifold in type IIB string theory.\footnote{Note that this is distinct from the case of a single brane descending a warped throat along both radial and angular coordinates, as in the first example of multiple-field DBI \cite{Easson08}.} First considered by \cite{Cai09,Cai09-1,Pi11} to investigate the effect of multiple sound horizons on equilateral type non-Gaussianity, the corresponding expression for $P_{I}$ is

      \be P_{I}=\frac{1}{f^{(I)}}\left(1-\sqrt{1-2f^{(I)}X_{I}}\right),\label{eq:Lagrangian-DBI}\ee

\noindent where $f^{(I)}$ parameterises the warped brane tension of throat $I$ and is a function of $\phi_{I}$ only. The corresponding homogeneous equations of motion are given by 

      \begin{align}	
      \dot{\phi}_{I}&=-2c^{(I)}H_{,I},\label{eq:field-dbi}\\[15pt]
      3H^{2}&=V-\sum_{I}\frac{1}{f^{(I)}}\left(1-\frac{1}{c^{(I)}}\right).\label{eq:friedman-dbi} 
      \end{align}
      
\noindent Here $c^{(I)}$ is defined as the sound speed associated with the field $I$

      \be c^{(I)}=\sqrt{\frac{P_{I,X_{I}}}{\rho_{I,X_{I}}}}=\frac{1}{\sqrt{1+4f^{(I)}H_{,I}^{2}}},\label{eq:sound-speed}\ee
      
\noindent and we have used (\ref{eq:field-dbi}) to eliminate $X_{I}$ such that $f^{(I)}$ remains a function of $\phi_{I}$ whilst $H$, $c^{(I)}$, $V$ and $\dot{\phi_{I}}$ generally depend on the collection of fields $\phi$. Notice that the above equations of motion are written in Hamilton-Jacobi form \cite{Salopek90,Kinney97} in which the Hubble parameter is written as a function of the scalar fields, taking precedence over the potential. This is more suited to the case of non-trivial sound speeds and will allow us to consider departures from slow variation. To define slow variation we introduce the following parameters
	    
      \begin{align}
      \epsilon&=-\frac{\dot{H}}{H^{2}}=\sum_{I}\epsilon^{(I)}=\sum_{I}2c^{(I)}\left(\frac{H_{,I}}{H}\right)^{2},\nonumber\\[10pt]
      \eta^{(I)}&=\sum_{J}\eta^{(IJ)}=\sum_{J}2c^{(J)}\frac{H_{,J}}{H_{,I}}\frac{H_{,IJ}}{H},\nonumber\\[10pt]
      s^{(I)}&=-\frac{\dot{c}^{(I)}}{Hc^{(I)}}=\sum_{J}s^{(IJ)}=\sum_{J}2c_{,J}^{(I)}\frac{c^{(J)}}{c^{(I)}}\frac{H_{,J}}{H},\label{eq:slow_roll_params}	
      \end{align}

\noindent where we require $\epsilon<1$ for inflation. Slow variation is  defined as $\epsilon^{(I)},\eta^{(I)},s^{(I)}\ll1$ and we shall state explicitly when this additional restriction is required. 

\subsection{Perturbations and non-Gaussianity\label{sec:model_pert}}

We characterise the scalar degree of freedom in the primordial density perturbations by introducing the primordial curvature perturbation on uniform density hypersurfaces $\zeta(t,x^{i})$ (see \cite{Malik01,Malik09} for explicit definitions). The two and three-point correlation functions then define the power spectrum $P_{\zeta}$  and bispectrum $B_{\zeta}$ respectively   

	\begin{align}
	\langle\zeta_{\mathbf{k_{1}}}\zeta_{\mathbf{k_{2}}}\rangle &= (2\pi)^{3} P_{\zeta}(k_{1}) \, \delta^{3}(\mathbf{k_{1}}+\mathbf{k_{2}}),\\[10pt]     
	\langle \zeta_{\mathbf{k_{1}}}\zeta_{\mathbf{k_{2}}}\zeta_{\mathbf{k_{3}}} \rangle &= 			(2\pi)^{3}B_{\zeta}(k_{1},k_{2},k_{3})\delta^{3}(\mathbf{k_{1}}+\mathbf{k_{2}}+\mathbf{k_{3}}),\label{eq:three-point-function}     
      \end{align}	    
	   
\noindent where $\zeta_{\mathbf{k}}$ is the Fourier transform of $\zeta$, $\mathbf{k_{i}}$ are comoving wavevectors and $\delta^{3}$ is the three dimensional Dirac delta function.  For a Gaussian $\zeta$ the two-point function completely defines the statistics of the field. Signatures of non-Gaussianity are encoded in the connected contributions to higher order correlators, such as the three-point function. To parameterise the deviation from Gaussianity we introduce the $k$-dependent non-linearity parameter\footnote{Here we adopt the sign conventions of \cite{Byrnes09} for ease of comparison. See \cite{Wands10} for a summary of the various conventions in the literature.} $\fnl$, given by the ratio of the bispectrum to a combination of power spectra
	    
      \be \frac{6}{5}\fnl(k_{1},k_{2},k_{3})=\frac{B_{\zeta}(k_{1},k_{2},k_{3})}{P_{\zeta}(k_{1})P_{\zeta}(k_{2}) + P_{\zeta}(k_{1})P_{\zeta}(k_{3}) + P_{\zeta}(k_{2})P_{\zeta}(k_{3})}.\ee

\noindent By assuming a scale invariant dimensionless power spectrum $\mathcal{P}_{\zeta}=\frac{k^{3}}{2\pi^{2}}P_{\zeta}(k)$ the above can be written as
	      
      \be  \frac{6}{5}\fnl(k_{1},k_{2},k_{3})=\frac{\prod_{i}k_{i}^{3}}{\sum_{i}k_{i}^{3}}\frac{B_{\zeta}(k_{1},k_{2},k_{3})}{4\pi^{4}\mathcal{P}_{\zeta}^{2}}.\label{eq:fnl}\ee

We now adopt the $\delta N$ formalism \cite{Starobinskivi85,Sasaki96,Sasaki98,Wands00,Lyth05-1} to evolve $\zeta$ on super-horizon scales using only the field fluctuations at horizon exit and the homogeneous field evolution thereafter. To facilitate this we make two restrictions on the background dynamics. First we demand that the sound speeds are comparable whilst observable scales exit during inflation, such that $c^{(I)} \simeq c_{\star}$ for all $I$ during this interval.\footnote{This need not necessarily be the case, as in \cite{Cai09-1,Pi11} for example.} Horizon exit\footnote{We will use the more succinct term `horizon' as opposed to `sound-horizon' since there should not be any ambiguity as, in this context, the sound-horizon is the only relevant scale.} therefore equates to evaluating a quantity when $c_{\star}k = a_{\star}H_{\star}$. To simplify the spectrum of field fluctuations we also assume slow variation during horizon exit.

Given these restrictions, we use the separate Universe approach \cite{Wands00,Sasaki96,Sasaki98,Rigopoulos03} to identify the curvature perturbation $\zeta$ with the difference in the number of e-folds between the perturbed ($N$) and homogeneous background ($N_{0}$) universes, evaluated between an initially flat hypersurface $t_{\star}$ (e.g. shortly after horizon exit) and a final uniform density hypersurface $t_{f}$ (e.g. early in the radiation dominated epoch). This allows us to calculate the relevant quantities (e.g. $\fnl$) at time $t_{f}$ given the field fluctuations at time $t_{\star}$ and the homogeneous field evolution between these times. For example, the dimensionless power spectrum can be written as \cite{Lyth05-1,Byrnes06}

      \be \mathcal{P}_{\zeta}=\sum_{I}N^{2}_{,I}\mathcal{P}_{\star},\label{eq:power-spectrum-deltaN}\ee	

\noindent where $N_{,I}$ is with respect to the field $I$ at horizon exit. Here we have defined the dimensionless power spectrum of scalar field fluctuations at horizon exit using the two point function

      \be \langle \delta\phi_{I\,\mathbf{k_{1}}} \delta \phi_{J\,\mathbf{k_{2}}}\rangle=(2\pi)^{3}\,\delta_{IJ}\frac{2\pi^{2}}{k_{1}^{3}}\mathcal{P}_{\star}\,\delta^{3}(\mathbf{k_{1}}+\mathbf{k_{2}}),\hspace{20pt}\mathcal{P}_{\star}=\left(\frac{H_{\star}}{2\pi}\right)^{2},\label{eq:two-point-fields}\ee

\noindent where we use slow variation at horizon exit and $\delta^{IJ}$ is the kronecker delta symbol. Similarly the three-point function is given by  
	    
      \begin{align}
      \langle \zeta_{\mathbf{k_{1}}}\zeta_{\mathbf{k_{2}}}\zeta_{\mathbf{k_{3}}} \rangle =  \sum_{IJK}&N_{,I}N_{,J}N_{,K}\langle\delta\phi_{I\,\mathbf{k_{1}}}\delta\phi_{J\,\mathbf{k_{2}}}\delta\phi_{K\,\mathbf{k_{3}}}\rangle \, +  \nonumber \\[5pt]  
      &\biggl(\,\frac{1}{2}\sum_{IJKL}N_{,I}N_{,J}N_{,KL}\langle\delta\phi_{I\,\mathbf{k_{1}}}\delta\phi_{J\,\mathbf{k_{2}}}(\delta\phi_{K}\star\delta\phi_{L})_{\mathbf{k_{3}}}\rangle + 2\,\mathrm{perms}\,\biggr)\label{eq:three-point-function-deltaN},
      \end{align}

\noindent where in this case $\star$ denotes a convolution and `perms' denotes cyclic permutations over the momenta. Neglecting the connected part of the four-point function and using Wick's theorem to rewrite the four-point functions as products of two-point functions, the latter term can be written as \cite{Lyth05-1,Byrnes06} 
	    
      \begin{align}
      \frac{1}{2}\sum_{IJKL}N_{,I}&N_{,J}N_{,KL}\langle\delta\phi_{I\,\mathbf{k_{1}}}\delta\phi_{J\,\mathbf{k_{2}}}(\delta\phi_{K}\star\delta\phi_{L})_{\mathbf{k_{3}}}\rangle + 2\,\mathrm{perms}=\nonumber \\[10pt]&(2\pi)^{3}\,4\pi^{4}\mathcal{P}_{\star}^{2}\,\frac{\sum_{i}k_{i}^{3}}{\prod_{i}k_{i}^{3}}\sum_{IJ}N_{,I}N_{,J}N_{,IJ}\,\delta^{3}({\mathbf{k_{1}}+\mathbf{k_{2}}+\mathbf{k_{3}}}),
      \end{align}

\noindent such that the bispectrum becomes
	    
      \begin{align}
      B_{\zeta}(k_{1},k_{2},k_{3})=4\pi^{4}\mathcal{P}_{\zeta}^{2}\frac{\sum_{i}k_{i}^{3}}{\prod_{i}k_{i}^{3}}\left(\frac{6}{5}\fnl^{(3)}(k_{1},k_{2},k_{3})+\frac{6}{5}\fnl^{(4)}\right).\label{eq:bispectrum}
      \end{align}

\noindent Inspection of $\fnl(k_{1},k_{2},k_{3})=\fnl^{(3)}(k_{1},k_{2},k_{3}) + \fnl^{(4)}$ shows that there are two distinct contributions to the bispectrum. Adopting the notation of \cite{Vernizzi06}, the $k$-independent parameter\footnote{For scale dependent cases see \cite{Byrnes10a,Byrnes10b}.} $\fnl^{(4)}$  is due to non-linear behaviour in $\zeta$ on super-horizon scales and is referred to as the local contribution, given by

      \be \frac{6}{5}\fnl^{(4)}=\frac{\sum_{IJ}N_{,I}N_{,J}N_{,IJ}}{(\sum_{K}N_{,K}^{2})^{2}}.\label{eq:fnl4}\ee

\noindent The result of our previous work was to provide an analytic expression for this parameter in the subset of cases described by a sum-separable Hubble parameter, in which the derivatives $N_{,I}$ and $N_{,IJ}$ can be fully evaluated. We neglected the contribution from the $k$-dependent parameter $\fnl^{(3)}(k_{1},k_{2},k_{3})$ however, which is due to the intrinsic non-Gaussianity of the $\delta\phi^{I}$, produced by quantum field interactions on sub-horizon scales. 

The aim of this paper then is to explicitly calculate the equilateral contribution and so arrive at the total expression for the bispectrum. Inspection of (\ref{eq:three-point-function-deltaN}) shows that this requires two distinct steps.  In the following section we use the path integral method to first calculate the three point function of field fluctuations at horizon crossing. Thereafter we use the $\delta N$ formalism to find the equilateral non-linearity parameter of the curvature perturbation in this scenario. 

\section{The equilateral contribution\label{sec:three_point}}

In this section we use a standard prescription to calculate the three point function of field fluctuations at horizon crossing, analogous to calculations in \cite{Seery05,Seery05b,Langlois08b,Arroja08}. We begin by presenting the third order action of field fluctuations for a more general scenario in the spatially flat gauge, before restricting ourselves to slow variation and small sound speeds around horizon exit in the multiple-DBI case. Thereafter, we use this result and the path integral formalism to find the three point function of field fluctuations at horizon exit and in turn an expression for $\fnl^{(3)}$.

\subsection{The third order action\label{sec:three_point_third_order_action}}

\noindent To calculate the three point function of field fluctuations we begin with the general action (\ref{eq:action-general}).
%
%
\noindent For a spatially flat Friedman-Robertson-Walker (FRW) Universe, the background equations of motion are given by

    \begin{align} 
    3H^{2}&=\sum_{I}\left[\dot{\phi}_{I}^{2}P_{I,X_{I}} - P_{I}\right]+V,\label{eq:friedmann}\\[10pt]
    \dot{H}&=-\frac{1}{2}\sum_{I}\dot{\phi}_{I}^{2}P_{I,X_{I}},\label{eq:field} 
    \end{align}

\noindent which here we write in the conventional form, as opposed to the Hamilton-Jacobi form in section~\ref{sec:model_homo}. The Klein-Gordon equation, which is not independent of (\ref{eq:friedmann}) and (\ref{eq:field}), is given by

    \be P_{I,I}=3HP_{I,X_{I}}\dot{\phi}_{I}+\dot{P}_{I,X_{I}}\dot{\phi}_{I}+P_{I,X_{I}}\ddot{\phi}_{I}+V_{,I}.\label{eq:klein-gordan} \ee

\noindent Progressing to perturbations about the homogeneous background, we construct the third order action by recasting (\ref{eq:action-general}) using the Arnowitt-Deser-Misner (ADM) formalism \cite{Arnowitt62}. This will be useful since the lapse function $N$ and shift vector $N^{i}$ become Lagrange multipliers under variation. This, along with an appropriate choice of gauge, will simplify the task of isolating the physical degrees of freedom when we consider perturbations. Until then, however, we stress that the equations remain exact with no choice of gauge. The ADM metric is given by

    \be ds^{2}=-N^{2}dt^{2}+h_{ij}\left(dx^{i}+N^{i}dt\right)\left(dx^{j}+N^{j}dt\right), \ee

\noindent where $h_{ij}$ is the spatial 3-metric. In terms of this metric, the action (\ref{eq:action-general}) and kinetic term become

    \begin{align} 	
    S&=\frac{1}{2} \int d^{4}x\sqrt{h}\Bigl[NR^{(3)}+NK_{ij}K^{ij}-NK^{2}+2N\sum_{I}P_{I}-2NV\Bigr],\label{eq:action-ADM}\\[10pt]
    X_{I}&=\frac{1}{2N^{2}}\left(\dot{\phi}_{I} - N^{i}\phi_{I,i}\right)^{2}-\frac{1}{2}\phi_{I,i}\phi_{I}^{,i},\label{eq:kinetic-ADM}
    \end{align}

\noindent where $K=K^{i}_{i}$, $R^{(3)}$ is the three dimensional Ricci scalar and indices are raised and lowered using the spatial metric. $K_{ij}$ is the extrinsic curvature, given by

    \be K_{ij}=\frac{1}{2N}\left(N_{i|j}+N_{j|i}-\dot{h}_{ij}\right), \ee

\noindent  where $_{|i}$ denotes the covariant derivative with respect to the spatial metric. To derive the energy and momentum constraint equations we vary the ADM action (\ref{eq:action-ADM}) with respect to the lapse function $N$ and shift vector $N^{i}$ respectively

    \be R^{(3)}-K_{ij}K^{ij}+K^{2}-2\sum_{I}\left[\frac{P_{I,X_{I}}}{N^{2}}v_{I}^{2}-P_{I}\right]+V=0,\label{eq:energy-constraint} \ee

    \be K_{|i}-K^{j}_{i|j}-\sum_{I}\frac{P_{I,X_{I}}}{N}v_{I}\phi_{I,i}=0,\label{eq:momentum-constraint} \ee

\noindent where for notational convenience we have introduced $v_{I}=\dot{\phi}_{I}-N^{i}\phi_{I,i}$. To solve the energy and momentum constraints we consider a first order\footnote{Note that it suffices to consider a first order expansion in the energy and momentum constraints since higher order contributions vanish on substitution into the action.} expansion of the inhomogeneous quantities about a spatially flat FRW background
    
    \be \phi_{I}=\bar{\phi_{I}}+\delta\phi_{I},\nonumber \ee
    \be N=1+\alpha,\hspace{20pt}N_{i}=\beta_{|i},\nonumber \ee
    \be h_{ij}=a^{2}\left((1-2\psi)\delta_{ij}+2E_{|ij}\right), \ee

\noindent where we consider scalar perturbations only and $\delta_{ij}$ is the Kronecker delta symbol. This presents $n+4$ scalar degrees of freedom: $\delta\phi_{I},\alpha,\beta,\psi$ and $E$. We can eliminate two degrees of freedom by adopting the spatially flat gauge, whereby $\psi=0$ and $E=0$, such that

    \be \phi_{I}=\bar{\phi_{I}}+\delta\phi_{I},\nonumber \ee  
    \be N=1+\alpha,\hspace{20pt}N_{i}=\beta_{|i},\nonumber \ee 
    \be h_{ij}=a^{2}\delta_{ij}, \ee

\noindent leaving $n+2$ scalar degrees of freedom. Note that for notational convenience we drop the overbar on homogeneous quantities for the remainder of this section. To eliminate two further degrees of freedom we substitute the above into the constraint equations (\ref{eq:energy-constraint}) and (\ref{eq:momentum-constraint}), giving algebraic equations for $\alpha$ and $\beta^{|i}_{\;\;|i}$

    \be \alpha=\sum_{I}\frac{P_{I,X_{I}}\dot{\phi}_{I}}{2H}\delta\phi_{I},\label{eq:energy-constraint-results} \ee
    
    \begin{align}
    \beta^{|i}_{\;\;|i}&=-\frac{1}{2H}\sum_{I}\Biggl[P_{I,IX_{I}}\dot{\phi}_{I}^{2}\delta\phi_{I}-P_{I,I}\delta\phi_{I}+V_{,I}\delta\phi_{I}   +\left(P_{I,X_{I}}+P_{I,X_{I}X_{I}}\dot{\phi}_{I}^{2}\right)\dot{\phi}_{I}\delta\dot{\phi}_{I}+\Biggr.\nonumber\\
    &\;\Biggl.+\left(3H^{2}+\frac{1}{2}\sum_{J}\left[P_{J,X_{J}}+P_{J,X_{J}X_{J}}\dot{\phi}_{J}^{2}\right]\dot{\phi}_{J}^{2}\right)\frac{P_{I,X_{I}}\dot{\phi}_{I}}{H}\delta\phi_{I}\Biggr.\label{eq:momentum-constraint-results}].
    \end{align}

\noindent The above can be substituted back into the action (\ref{eq:action-ADM}) expanded to the desired order and, after removing total derivatives and using the background equations of motion, yields the perturbed action in terms of the $n$ physical degree of freedom $\delta\phi_{I}$. For simplicity we begin with the second order action

    \begin{align}
    S_{(2)}&=\frac{1}{2}\int d^{4}x\,a^{3}\sum_{I}\Biggl[\left(P_{X_{I}}+P_{X_{I}X_{I}}\dot{\phi}_{I}^{2}\right)\delta\dot{\phi}_{I}^{2}-P_{X_{I}}\delta\phi_{I,i}\delta\phi_{I}^{,i}\Biggr. \nonumber\\
    &\Biggr.-\sum_{J}M_{IJ}\delta\phi_{I}\delta\phi_{J}+\sum_{J}\left[2P_{I,IX_{I}}\dot{\phi}_{I}\delta_{IJ}-\frac{1}{H}P_{I,X_{I}X_{I}}P_{J,X_{J}}\dot{\phi}_{I}^{3}\dot{\phi}_{J}\right]\delta\dot{\phi}_{I}\delta\phi_{J}\Biggr],\label{eq:action-second}
    \end{align}

\noindent where the effective mass matrix is given by:

    \begin{align}
    M_{IJ}&=-P_{I,II}\delta_{IJ}+V_{,IJ}+\frac{1}{2H}\left(P_{I,IX_{I}}P_{J,X_{J}}\dot{\phi}_{I}^{2}\dot{\phi}_{J}+P_{J,JX_{J}}P_{I,X_{I}}\dot{\phi}_{J}^{2}\dot{\phi}_{I}\right)\nonumber\\[10pt]
    &-\frac{1}{4H^{2}}\sum_{K}P_{I,X_{I}}P_{J,X_{J}}P_{K,X_{K},X_{K}}\dot{\phi}_{I}\dot{\phi}_{J}\dot{\phi}_{K}^{4}-\frac{1}{a^{3}}\frac{d}{dt}
    \left(\frac{a^{3}}{H}P_{I,X_{I}}P_{J,X_{J}}\dot{\phi}_{I}\dot{\phi}_{J}\right),\label{eq:action-second}
    \end{align}

\noindent which can be used to evaluate the two point function of field fluctuations, as in (\ref{eq:two-point-fields}), using the standard prescription. We then find, after some lengthy calculations, the corresponding third order action (see \cite{Seery05b,Langlois08b,Arroja08} for analogous calculations)

    \begin{align}
    S_{(3)}&=\int dtd^{3}x\,a^{3}\Biggl[\left(3H^{2}\alpha^{2}+2H\alpha\beta^{|i}_{\;\;|i}+\frac{1}{2}\left(\beta^{|i}_{\;\;|i}\beta^{|j}_{\;\;|j}-\beta_{|ij}\beta^{|ij}\right)\right)\alpha\Biggr.\nonumber\\[10pt]
    &+\sum_{I}\Biggl[\left(-\frac{1}{2}\dot{\phi}_{I}^{2}\alpha^{3}+\dot{\phi}_{I}\alpha^{2}\delta\dot{\phi}_{I}+\dot{\phi}_{I}\alpha\beta^{|i}\delta\phi_{I|i}-\frac{1}{2}\alpha\delta\dot{\phi}_{I}^{2}-\left(\beta^{|i}\delta\dot{\phi}_{I}+\frac{1}{2}\alpha\delta\phi_{I}^{|i}\right)\delta\phi_{I|i}\right)P_{I,X_{I}}\Biggr.\nonumber\\[10pt]
    &+\left(\dot{\phi}_{I}^{2}\alpha^{2}-\frac{3}{2}\dot{\phi}_{I}\alpha\delta\dot{\phi}_{I}+\frac{1}{2}\delta\dot{\phi}_{I}^{2}-\left(\dot{\phi}_{I}\beta^{|i}+\frac{1}{2}\delta\phi_{I}^{|i}\right)\delta\phi_{I|i}\right)P_{I,X_{I}X_{I}}\tilde{X}_{I}+\frac{1}{2}P_{I,IX_{I}X_{I}}\tilde{X}_{I}^{2}\delta\phi_{I}\nonumber\\[10pt]
    &+\left(\frac{1}{2}\dot{\phi}_{I}^{2}\alpha^{2}-\dot{\phi}_{I}\alpha\delta\dot{\phi}_{I}+\frac{1}{2}\delta\dot{\phi}_{I}^{2}-\left(\dot{\phi}_{I}\beta^{|i}+\frac{1}{2}\delta\phi_{I}^{|i}\right)\delta\phi_{I|i} \right)P_{I,IX_{I}}\delta\phi_{I}+\frac{1}{2}P_{I,IIX_{I}}\tilde{X}_{I}\delta\phi_{I}^{2}\nonumber\\[10pt]
    &+\,\frac{1}{2}P_{I,II}\alpha\delta\phi_{I}^{2}+\frac{1}{6}P_{I,III}\delta\phi_{I}^{3}-\sum_{J}\frac{1}{2}V_{,IJ}\alpha\delta\phi_{I}\delta\phi_{J}-\sum_{J}\sum_{K}\frac{1}{6}V_{,IJK}\delta\phi_{I}\delta\phi_{J}\delta\phi_{K}\nonumber\\
    &\,\Biggl.\Biggl.\,+\frac{1}{6}P_{I,X_{I}X_{I}X_{I}}\tilde{X}_{I}^{3}\Biggr]\Biggr],\label{eq:action-third}
    \end{align}

\noindent where for notational convenience we have introduced $\tilde{X}_{I}=\dot{\phi}_{I}\left(\delta\dot{\phi}_{I}-\dot{\phi}_{I}\alpha\right)$ and we note again that only the first order energy (\ref{eq:energy-constraint-results}) and momentum (\ref{eq:momentum-constraint-results}) constraints are required. The above results are consistent with an analogous calculation by \cite{Arroja08}, who consider a slightly more general action where $P$ in Eq.~(\ref{eq:action-second}) is a function of the kinetic functions $X_{IJ}=-\frac{1}{2}g^{\mu\nu}\phi_{I,\mu}\phi_{J,\nu}$ and the scalar fields $\phi=\{ \phi_{1},\phi_{2},...,\phi_{n}\}$. Note that where we use $a^{-2}\delta^{ij}$ to raise spatial indices, \cite{Arroja08} use $\delta^{ij}$ only, which accounts for the additional factors of $a^{-2}$ and $a^{-4}$ in the latter's results. The two sets of expressions are identical when the metric is written explicitly.

We now consider the leading contributions to the action (\ref{eq:action-third}) in slow variation and small sound speeds, since the dominant contribution in the following path integrals will be around horizon exit. Neglecting the purely gravitational part of the action and following the arguments regarding the more general versions of slow variation in \cite{Langlois08b,Arroja08} we find

    \begin{align}
    S_{(3)}\simeq\int dtd^{3}x\,a^{3}\sum_{I}&\left[\left(\frac{1}{2}\dot{\phi}_{I}P_{I,X_{I}X_{I}}+\frac{1}{3}\dot{\phi}_{I}X_{I}P_{I,X_{I},X_{I},X_{I}}\right)\delta\dot{\phi}_{I}^{3}\right.\nonumber\\[10pt]
    &\,\,\,\,\left.-\frac{1}{2}\dot{\phi}_{I}P_{I,X_{I}X_{I}}\delta\dot{\phi}_{I}\delta\phi_{I}^{|i}\delta\phi_{I|i}\right].\label{eq:action-third-before}
    \end{align}

Consider now the multiple-DBI scenario, described by the Lagrangian (\ref{eq:Lagrangian-DBI}), which we repeat for reference

    \be P_{I}=\frac{1}{f^{(I)}}\left(1-\sqrt{1-2f^{(I)}X_{I}}\right).\ee

\noindent It is then straight forward to compute the following derivatives

    \be P_{I,X_{I}}=\frac{1}{c_{I}},\hspace{25pt}P_{I,X_{I}X_{I}}=\frac{f_{I}}{c_{I}^{3}},\hspace{25pt}P_{I,X_{I}X_{I},X_{I}}=\frac{3f_{I}^{2}}{c_{I}^{5}}. \ee

\noindent Upon substitution into (\ref{eq:action-third-before}) and keeping only terms at leading order small sound speeds,  we arrive at 
    
    \begin{align}
    S_{(3)}\simeq\int dtd^{3}x\,a^{3}\sum_{I}\Biggl[\frac{1}{2}\frac{1}{\dot{\phi}_{I}c_{I}^{5}}\delta\dot{\phi}_{I}^{3}-\frac{1}{2}\frac{1}{\dot{\phi}_{I}c_{I}^{3}}\delta\dot{\phi}_{I}\delta\phi_{I}^{|i}\delta\phi_{I|i}\Biggr].\label{eq:action-third-leading-DBI}
    \end{align}

\noindent This is the third-order action for field fluctuations in the multi-DBI scenario to leading order in slow variation and small sound speeds, which is justified around horizon exit.   

\subsection{The path integral formalism\label{sec:three_point_path_integral}}

\noindent With the action (\ref{eq:action-third-leading-DBI}) we are now in a position to calculate the three point function of field fluctuations at horizon crossing to leading order in slow variation and small sound speeds. To this end we adopt the path integral technique and, for brevity, we refer to \cite{Seery05b} for a clear and detailed description of this method. We first require some standard results, the first being the propagator and its time derivative. The following is easily obtained from the second order theory (\ref{eq:action-second}) assuming slow variation at horizon exit, in exactly the same way as the power spectrum (\ref{eq:two-point-fields}) 

\begin{align} 
\langle\delta\phi_{I\,\mathbf{k_{1}}}(\tau_{1})&\delta\phi_{J\,\mathbf{k_{2}}}(\tau_{2})\rangle=(2\pi)^{3}\frac{H^{2}}{2k_{1}^{3}}(1+ic_{I}k_{1}\tau_{1})\,\times\nonumber\\[15pt]
&(1-ic_{I}k_{1}\tau_{2})e^{-ik_{1}c_{I}(\tau_{1}-\tau_{2})}\delta_{IJ}\delta^{(3)}(\mathbf{k_{1}}+\mathbf{k_{2}}+\mathbf{k_{3}}),\label{eq:propagator} 
 \end{align}

\begin{align} 
\frac{d}{d\tau_{2}}\langle\delta\phi_{I\,\mathbf{k_{1}}}(\tau_{1})&\delta\phi_{J\,\mathbf{k_{2}}}(\tau_{2})\rangle=(2\pi)^{3}\frac{H^{2}c_{I}^{2}}{2k_{1}}\,\times\nonumber\\[15pt]
&\tau_{2}(1+ic_{I}k_{1}\tau_{1})e^{-ik_{1}c_{I}(\tau_{1}-\tau_{2})}\delta_{IJ}\delta^{(3)}(\mathbf{k_{1}}+\mathbf{k_{2}}+\mathbf{k_{3}}).
 \end{align}

\noindent where $\tau$ represents conformal time. Indeed, by considering equal times $\tau_{1}=\tau_{2}$ in the super-horizon limit $|c_{I}k_{1}\tau|\ll 1$, the above yields exactly the definition of the dimensionless power spectrum (\ref{eq:two-point-fields}). In addition we require the following time integrals, which can be obtained by choosing the appropriate contour in the complex plane (analogous integrals appear in \cite{Langlois08})

\be \int_{-\infty}^{0}e^{iKc_{I}\tau}d\tau=-\frac{i}{Kc_{I}},\hspace{15pt}\int_{-\infty}^{0}\tau e^{iKc_{I}\tau}d\tau=\frac{1}{(Kc_{I})^{2}},\hspace{15pt}\int_{-\infty}^{0}\tau^{2}e^{iKc_{I}\tau}d\tau=\frac{2i}{(Kc_{I})^{3}}. \ee

\noindent Given the above we are now in a position to use the standard prescription, as in \cite{Seery05b}, to find the following contributions to $\langle\delta\phi_{I\,\mathbf{k_{1}}}\delta\phi_{J\,\mathbf{k_{2}}}\delta\phi_{K\,\mathbf{k_{3}}}\rangle$ from the first and second terms in (\ref{eq:action-third-leading-DBI}) respectively 

\be (2\pi)^{3}\frac{6}{4}\frac{H^{4}}{\sqrt{2\epsilon_{I}c_{I}}}\frac{1}{c_{I}^{2}}\frac{1}{\prod_{i}k_{i}^{3}}\frac{k_{1}^{2}k_{2}^{2}k_{3}^{2}}{K^{3}}\delta_{IJ}\delta_{IK}\delta^{(3)}(\mathbf{k_{1}}+\mathbf{k_{2}}+\mathbf{k_{3}}), \ee

\be -(2\pi)^{3}\frac{1}{4}\frac{H^{4}}{\sqrt{2\epsilon_{I}c_{I}}}\frac{1}{c_{I}^{2}}\frac{k_{1}^{2}(\mathbf{k_{2}}\cdot\mathbf{k_{3}})}{\prod_{i}k_{i}^{3}}\left(\frac{1}{K}+\frac{(k_{2}+k_{3})}{K^{2}}+\frac{2k_{2}k_{3}}{K^{3}}\right)\delta_{IJ}\delta_{IK}\delta^{(3)}(\mathbf{k_{1}}+\mathbf{k_{2}}+\mathbf{k_{3}}), \ee

\noindent where terms in the above are to be evaluated at horizon crossing, since this is the dominant contribution to the relevant time integrals. Note that there is a sign ambiguity in the above from using (\ref{eq:slow_roll_params}) to write $H_{,I}$ in terms of $\epsilon^{(I)}$. The above are valid provided we assume $\dot{\phi_{I}}<0$, which will be the case in the scenarios we consider. Finally then, we sum these contributions to arrive at the three point function of field fluctuations at horizon exit, to leading order in slow variation and small sound speeds 

\begin{align}
\langle\delta\phi_{I\,\mathbf{k_{1}}}\delta\phi_{J\,\mathbf{k_{2}}}\delta\phi_{K\,\mathbf{k_{3}}}\rangle=(2\pi)^{3}\frac{1}{4}\frac{H^{4}}{c_{I}^2}\frac{1}{\sqrt{2\epsilon_{I}c_{I}}}\frac{\Lambda(k_{1},k_{2},k_{3})}{\prod_{i}k_{i}^{3}}\delta_{IJ}\delta_{JK}\delta^{(3)}(\mathbf{k_{1}}+\mathbf{k_{2}}+\mathbf{k_{3}}),\label{eq:fields-result}
\end{align}

\noindent where the $k$-dependent parameter $\Lambda$ is given by

\be \Lambda(k_{1},k_{2},k_{3})=\frac{6k_{1}^{2}k_{2}^{2}k_{3}^{2}}{K^{3}}-\left[k_{1}^{2}(\mathbf{k_{2}}\cdot\mathbf{k_{3}})\left(\frac{1}{K}+\frac{(k_{2}+k_{3})}{K^{2}}+\frac{2k_{2}k_{3}}{K^{3}}\right)+\textrm{perms}\right]. \ee

\noindent It is then trivial to check that the above expressions recover the single field result \cite{Chen05a}.  

\subsection{The equilateral non-linearity parameter\label{sec:full_three_point_fnl3}}

\noindent Given the expression for the three point function of the field fluctuations at horizon exit (\ref{eq:fields-result}), it is then straight forward to find the corresponding contribution to the three point function of $\zeta$ using the $\delta N$ formalism. To this end, we substitute the result (\ref{eq:fields-result}) into the first term in (\ref{eq:three-point-function-deltaN}) and consider the two field scenario with fields $\phi$ and $\chi$

\be \langle \zeta_{\mathbf{k_{1}}}\zeta_{\mathbf{k_{2}}}\zeta_{\mathbf{k_{3}}} \rangle _{\textrm{eq}} = (2\pi)^{3}\frac{1}{4}\frac{H_{\star}^{4}}{c_{\star}^{2}}\frac{\Lambda(k_{1},k_{2},k_{3})}{\prod_{i}k_{i}^{3}}\left(\frac{N_{,\phi}^{3}}{\sqrt{2\epsilon^{(\phi)}_{\star}c_{\star}}}+\frac{N_{,\chi}^{3}}{\sqrt{2\epsilon^{(\chi)}_{\star}c_{\star}}}\right)\delta^{(3)}(\mathbf{k_{1}}+\mathbf{k_{2}}+\mathbf{k_{3}}), \ee

\noindent where the subscript `eq' denotes that we are considering only the equilateral contribution. Comparison of the above with the definition of the bispectrum (\ref{eq:three-point-function}) gives,

\begin{align}
B_{\zeta}(k_{1},k_{2},k_{3})_{\textrm{eq}}&=\frac{1}{4}\frac{H_{\star}^{4}}{c_{\star}^{2}}\frac{\Lambda(k_{1},k_{2},k_{3})}{\prod_{i}k_{i}^{3}}\left(\frac{N_{,\phi}^{3}}{\sqrt{2\epsilon^{(\phi)}_{\star}c_{\star}}}+\frac{N_{,\chi}^{3}}{\sqrt{2\epsilon^{(\chi)}_{\star}c_{\star}}}\right)\nonumber\\[15pt]
&=4\pi^{4}\mathcal{P}_{\zeta}^{2}\frac{\sum_{i}k_{i}^{3}}{\prod_{i}k_{i}^{3}}\left[\frac{1}{c_{\star}^{2}}\frac{\Lambda(k_{1},k_{2},k_{3})}{\sum_{i}k_{i}^{3}}\frac{\left(\frac{N_{,\phi}^{3}}{\sqrt{2\epsilon^{(\phi)}_{\star}c_{\star}}}+\frac{N_{,\chi}^{3}}{\sqrt{2\epsilon^{(\chi)}_{\star}c_{\star}}}\right)}{(N_{,\phi}^{2}+N_{,\chi}^{2})^{2}} \right],\label{eq:result-bispectrum-eq}
\end{align}

\noindent where we have used (\ref{eq:power-spectrum-deltaN}) to replace $\mathcal{P}_{\star}$ with $\mathcal{P}_{\zeta}$. Finally then, the terms in parenthesis can be associated with the non-linearity parameter $\fnl^{(3)}$ by inspection of (\ref{eq:bispectrum})

\be \fnl^{(3)}(k_{1},k_{2},k_{3}) = \frac{5}{6}\frac{1}{c_{\star}^{2}}\frac{\Lambda(k_{1},k_{2},k_{3})}{\sum_{i}k_{i}^{3}}\frac{\left(\frac{N_{,\phi}^{3}}{\sqrt{2\epsilon^{(\phi)}_{\star}c_{\star}}}+\frac{N_{,\chi}^{3}}{\sqrt{2\epsilon^{(\chi)}_{\star}c_{\star}}}\right)}{(N_{,\phi}^{2}+N_{,\chi}^{2})^{2}}.\label{eq:fnl3}\ee

\noindent The above expression for $\fnl^{(3)}$ is the main result of this section and, in the absence of additional dynamical restrictions to evaluate the derivatives of $N$, cannot be developed further analytically. The $k$-dependence is unchanged compared to that of the single field scenario \cite{Chen05a}, such that this contribution does indeed peak in the equilateral limit $k_{1}=k_{2}=k_{3}$. By considering $\dot{\chi}\rightarrow 0$, such that $N_{,\chi}\rightarrow 0$ and $N_{,\phi}\rightarrow\frac{H}{\dot{\phi}}$, and using the background equations of motion (\ref{eq:field-dbi}), it can be shown that the final terms becomes equal to one, recovering the single field result.

We notice then that the above has the form of the single field result, which is precluded observationally by the strong $c_{\star}^{-2}$ dependence, modulated by an expression dependent on the background evolution after horizon exit. It seems possible then that, in some circumstances, this modulation may suppress the value of $\fnl^{(3)}$ to remain within observational bounds. Such modulation has been found in similar multiple-field DBI scenarios. For example, \cite{Langlois08b} use the adiabatic-entropy perturbation basis to obtain the single field result modulated by a $\cos^{2}\Theta$ term, where $\Theta$ depends on the background trajectory after horizon exit. This modulation has then been exploited in concrete examples to suppress the otherwise observationally precluded value of $\fnl^{(3)}$, as in \cite{Kidani12}. It is not \emph{a priori} obvious if this possible in our case by inspection of (\ref{eq:fnl3}) however, since the behaviour of this term is highly model dependent. We address this issue in the following section by considering scenarios in which the derivatives of $N$ can be evaluated analytically, as we did for the local contribution in \cite{Emery12}. 

\section{The total three point correlation function\label{sec:full_three_point}}

\noindent The expressions for $\fnl^{(3)}$ (\ref{eq:fnl3}) and $\fnl^{(4)}$ (\ref{eq:fnl4}) contain field derivatives of the number of e-folds $N$. Given the lack of a unique attractor in multiple-field scenarios however, additional restrictions to the background dynamics are required to further develop such expressions analytically. For example, in our previous work \cite{Emery12} we adopted the method of \cite{Byrnes09} and demanded a sum separable Hubble parameter\footnote{As opposed to a sum separable potential $V(\phi,\chi)=V(\phi)+V(\chi)$, as was used in the original application of this technique \cite{Vernizzi06}.}. Not only did this allow the violation of slow variation after horizon exit, it also suited the case of non-standard kinetic terms, since the dynamics are better described in the Hamilton-Jacobi formalism. Given this restriction, we exploited the resultant integral of motion to derive analytic expressions for the derivatives of $N$ and in turn $\fnl^{(4)}$ in the multiple-DBI case. In this section we apply those results to present an analogous expression for $\fnl^{(3)}$ which, to the best of our knowledge, is the first example of the application of the separable technique towards the equilateral contribution.

To make analytical progress we now restrict our attention to two-field models, with fields $\phi$ and $\chi$, that posses a sum separable Hubble parameter
     
      \be H(\phi,\chi)=H^{(\phi)}(\phi)+H^{(\chi)}(\chi),\label{eq:separable-Hubble}\ee
      
\noindent which leads to a number of simplifications. Inspection of (\ref{eq:sound-speed}) shows that the sound speed $c^{(I)}$ becomes a function of its respective field $\phi_{I}$ only, such that $c^{(\phi)}(\phi)$ and $c^{(\chi)}(\chi)$. Moreover, mixed derivatives of $H$ (i.e. $H_{,\phi\chi}$) become zero. The combination of the above reduces the number of relevant slow variation parameters
	
      \vspace{-3pt}
      \begin{align}
      \epsilon^{(\phi)}&=2c^{(\phi)}\left(\frac{H_{,\phi}}{H}\right)^{2},\hspace{35pt}
      \epsilon^{(\chi)}=2c^{(\chi)}\left(\frac{H_{,\chi}}{H}\right)^{2},\label{eq:epsilon-separable}\\[13pt]
      \eta^{(\phi)}&=2c^{(\phi)}\frac{H_{,\phi\phi}}{H},\hspace{53pt}
      \eta^{(\chi)}=2c^{(\chi)}\frac{H_{,\chi\chi}}{H},\label{eq:eta-separable}\\[13pt]
      s^{(\phi)}&=2c^{(\phi)}_{,\phi}\frac{H_{,\phi}}{H},\hspace{58pt}
      s^{(\chi)}=2c^{(\chi)}_{,\chi}\frac{H_{,\chi}}{H},\label{eq:s-separable}
      \end{align}
      \vspace{0pt}

\noindent where $\epsilon=\epsilon^{(\phi)}+\epsilon^{(\chi)}$ and we emphasise again that $\eta^{(I)}$ and $s^{(I)}$ can become much greater than one after horizon exit. Crucially though, the above assumption enables us to calculate the field derivatives of $N$ analytically by defining an integral of motion. Here we simply quote the results of our previous work \cite{Emery12}, where the full details of the calculation can be found. The derivatives of $N$ can be expressed in terms of slow variation parameters
      
      \be N_{,\phi_{\star}}=\frac{1}{\sqrt{2\epsilon^{(\phi)}_{\star}c_{\star}}}u,\hspace{30pt}N_{,\chi_{\star}}=\frac{1}{\sqrt{2\epsilon^{(\chi)}_{\star}c_{\star}}}v,\label{eq:first-derivatives}\ee

\noindent where, for brevity, we have introduced the following definitions 

      \be u=\frac{H^{(\phi)}_{\star}+Z_{f}}{H_{\star}},\hspace{30pt}
      v=\frac{H^{(\chi)}_{\star}-Z_{f}}{H_{\star}},\hspace{30pt}
      Z_{f}=\frac{H^{(\chi)}_{f}\epsilon^{(\phi)}_{f}-H^{(\phi)}_{f}\epsilon^{(\chi)}_{f}}{\epsilon_{f}}.\ee
			
\noindent On substitution of the above into the result (\ref{eq:fnl3}) , we arrive at the following expression for $\fnl^{(3)}$  

\be \fnl^{(3)}(k_{1},k_{2},k_{3})=\frac{5}{6}\frac{\Lambda(k_{1},k_{2},k_{3})}{\sum_{i}k_{i}^{3}}\frac{1}{c_{\star}^{2}}\frac{\left(\frac{u^{3}}{\epsilon_{\star}^{(\phi)^{2}}}+\frac{v^{3}}{\epsilon_{\star}^{(\chi)^{2}}}\right)}{\left(\frac{u^{2}}{\epsilon_{\star}^{(\phi)}}+\frac{v^{2}}{\epsilon_{\star}^{(\chi)}}\right)^{2}},\label{eq:fnl3-separable}\ee

\noindent which we emphasise is valid for the two-field DBI scenario assuming comparable small sound speeds and slow variation at horizon exit, in addition to a separable Hubble parameter. By setting $\dot{\chi}\rightarrow0$, we find $Z_{f}\rightarrow H_{f}^{(\chi)}$, $u\rightarrow1$ and $v\rightarrow0$ such that (\ref{eq:fnl3-separable}) again recovers the single field result \cite{Chen05a}. The above is analogous to the result (\ref{eq:fnl3}), in that we find the single field result modulated by a term dependent on the background evolution after horizon exit. The advantage here however, is that the behaviour of the modulation becomes more transparent. For example, we note that this term is approximately $\mathcal{O}(1)$ in slow variation at horizon exit. Moreover, since neither of the terms in the numerator are positive definite it is conceivable that, with a sufficient level of cancellation, the modulation term may suppress the otherwise prohibitively large contribution of $c_{\star}^{-2}$, providing an observationally viable value of $\fnl^{(3)}$ at the end of inflation. It remains to be seen if this is possible in a concrete setup however, which we intend to address in the following section.   

Alongside the above equilateral contribution (\ref{eq:fnl3-separable}), we must also consider the local signal. Here we can directly quote the analogous expression for  $\fnl^{(4)}$ from \cite{Emery12}
      
      \be \fnl^{(4)}=\frac{5}{6}\,\frac{\frac{u^{2}}{\epsilon^{(\phi)}_{\star}}\left(1-\frac{(\eta^{(\phi)}_{\star}+s^{(\phi)}_{\star})}
      {\epsilon^{(\phi)}_{\star}}u\right)+
      \frac{v^{2}}{\epsilon^{(\chi)}_{\star}}\left(1-\frac{(\eta^{(\chi)}_{\star}+s^{(\chi)}_{\star})}
      {\epsilon^{(\chi)}_{\star}}v\right)+
      2\left(\frac{u}{\epsilon^{(\phi)}_{\star}}-\frac{v}{\epsilon^{(\chi)}_{\star}}\right)^{2}\mathcal{A}}
      {\left(\frac{u^{2}}{\epsilon^{(\phi)}_{\star}}+\frac{v^{2}}{\epsilon^{(\chi)}_{\star}}\right)^{2}}. 
     \label{eq:fnl4-separable}\ee
      \vspace{5pt}
      
\noindent where the parameter $\mathcal{A}$ is defined as
      
      \be \mathcal{A}=-\frac{H_{f}^{2}}{H_{\star}^{2}}\frac{\epsilon^{(\phi)}_{f}\epsilon^{(\chi)}_{f}}{\epsilon_{f}}\left(\frac{1}{2}-\frac{\eta_{f}^{ss}}{\epsilon_{f}}-\frac{1}{2}\frac{s^{ss}_{f}}{\epsilon_{f}}\right),\label{eq:A}\ee
      
\noindent and we have introduced

	\be \eta^{ss}=\frac{\epsilon^{(\chi)}\eta^{(\phi)}+\epsilon^{(\phi)}\eta^{(\chi)}}{\epsilon},\hspace{10pt}s^{ss}=\frac{\epsilon^{(\chi)}s^{(\phi)}+\epsilon^{(\phi)}s^{(\chi)}}{\epsilon}. \ee
	
\noindent These additional parameters appear because the expression for $\fnl^{(4)}$ (\ref{eq:fnl4}) contains second derivatives of $N$, whilst $\fnl^{(3)}$ (\ref{eq:fnl3}) has only  first. Note that, whilst the terms preceding the parenthesis in the expression for $\mathcal{A}$ (\ref{eq:A}) are $\mathcal{O}(\epsilon_{\star})$, $\eta^{ss}_{f}$ and $s^{ss}_{f}$ can become much larger than unity, producing observable local type non-Gaussianity. We demonstrated this in a concrete model in \cite{Emery12} by considering inflation in the tip regions of two warped throats,  in which $s^{ss}_{f}$ becomes enhanced by the abrupt change of $c^{(\phi)}$ and $c^{(\chi)}$ at the end of inflation.

\noindent Equations (\ref{eq:fnl3-separable}) and (\ref{eq:fnl4-separable}) together provide the full expression for $f_{NL}$, and in turn the bispectrum (\ref{eq:bispectrum}), to leading order in slow variation and small, comparable sound speeds at horizon exit, given a sum separable Hubble parameter. We have noted that both contributions can, in principle at least, produce contributions to the bispectrum, rendering this one of the few explicit models capable of doing so (see \cite{RenauxPetel09,Kidani12} for alternatives). It remains to be seen if this is possible in practice however. Ideally, a consistency relation between the two contributions would elucidate this point further but, given the considerable  freedom within the model, finding a general relation has so far proved difficult. In the absence of a consistency relation therefore, it is useful to consider some specific models, as we did in \cite{Emery12} to study the local contribution produced by inflation in the tip regions of the throats. We leave this for the following section and here simply highlight that this scenario does indeed provide the potential mechanism to produce a mixed non-Gaussian signal.   

Before proceeding to specific scenarios, we first note that whilst the expressions (\ref{eq:fnl3-separable}) and (\ref{eq:fnl4-separable}) describe the production and evolution of non-Gaussianity \emph{during} inflation, these are not necessarily the final observed values. For completeness  we must track their evolution from the end of inflation until they are imprinted on the cosmic microwave background (CMB) at decoupling. Given our lack of knowledge of the early Universe however, this is not generally feasible. As such, recent work has considered whether such non-Gaussianity produced during inflation can indeed imprint upon the CMB \cite{Elliston11a,Elliston11b,Meyers11-a,Meyers11-b,Peterson11a,Watanabe11,Choi12,Mazumdar12}. Such study provides valuable clues as to what we can infer about inflationary dynamics from observations of non-Gaussianity and it would be interesting to include such considerations in this scenario. Here however,  we simply illustrate the potential production of mixed non-Gaussianity through multiple-field dynamics and small sound speeds during inflation. 

\section{Illustrative Examples\label{sec:models}}

In the previous section we demonstrated the possibility of producing mixed local and equilateral non-Gaussianity in the multiple-DBI scenario, through the expressions for $\fnl^{(3)}$ (\ref{eq:fnl3-separable}) and $\fnl^{(4)}$ (\ref{eq:fnl4-separable}). The aim of this section is to look more closely at the feasibility of producing such a signal by considering some concrete examples. We begin by revisiting the case of inflation in the tip regions of two warped throats, where previously we found enhanced local non-Gaussianity caused by the rapid increase in the sound speeds at the end of inflation. We find that the required modulation is insufficient to suppress the prohibitively large contribution of $c_{\star}^{-2}$ in (\ref{eq:fnl3-separable}) however, and leave further exploration of parameter space to future work. We then consider a phenomenologically similar but fully analytic model by choosing exponential warp factors, finding a regime in which observationally viable mixed non-Gaussianity is produced. Whilst not derived from string theory, this does provide a first step towards understanding the regimes in which this occurs, before progressing to more realistic setups.  

\subsection{Inflation in two cut-off throats\label{sec:models-cutoff}}

Previously, we studied the case of inflation in the tip regions of two warped throats and found enhanced local non-Gaussianity at the end of inflation, caused by a sudden increase in the sound speeds \cite{Emery12}. Here we revisit exactly this model and include the equilateral contribution, to study whether the background evolution is sufficient to suppress the contribution of $c_{\star}^{-2}$ in (\ref{eq:fnl3-separable}). We keep our discussion of the model brief, since full details can be found in \cite{Emery12}.

We choose a model whereby two probe D3 branes traverse two distinct warped throats glued to a compact Calabi-Yau in type IIB string theory \cite{Cai09,Cai09-1,Pi11}. As such, the warp factors are given by

      \be f^{(\phi)}=\frac{\lambda_{1}}{\phi^{4}}\left(1+\lambda_{2}\,\log\left(\frac{\phi}{\lambda_{3}}\right)\right),\label{eq:warp-factor}\ee

\noindent where $f^{(\chi)}$ is given by replacing $\phi\rightarrow\chi$ and for simplicity we have assumed the same warping in each throat. A full discussion of the physical significance of the $\lambda_{i}$ and the infrared singularity at $\phi=\lambda_{3}\,e^{-1/\lambda_{2}}$ can be found in \cite{Emery12}. We also choose a linear, separable Hubble parameter

      \be H(\phi,\chi)=\mathcal{H}^{(\phi)}\phi+\mathcal{H}^{(\chi)}\chi,\ee
      
\noindent such that $\eta^{(\phi)}\hspace{-3pt}=\eta^{(\chi)}\hspace{-3pt}=0$. Using (\ref{eq:sound-speed}), we arrive at the following expression for the sound speeds

\begin{figure}[t]
      \begin{center}
      \includegraphics[scale=0.5]{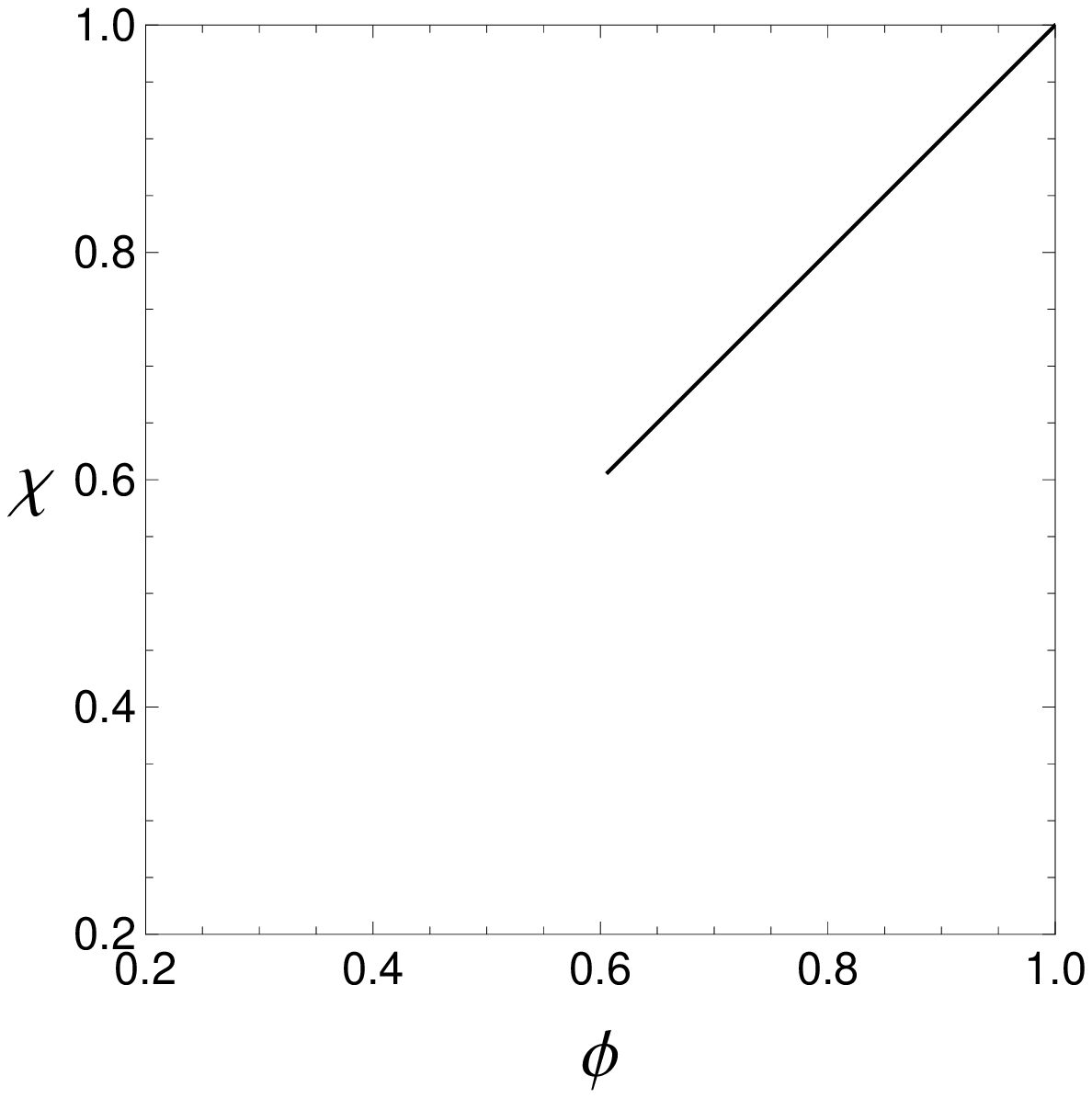}
      \includegraphics[scale=0.5]{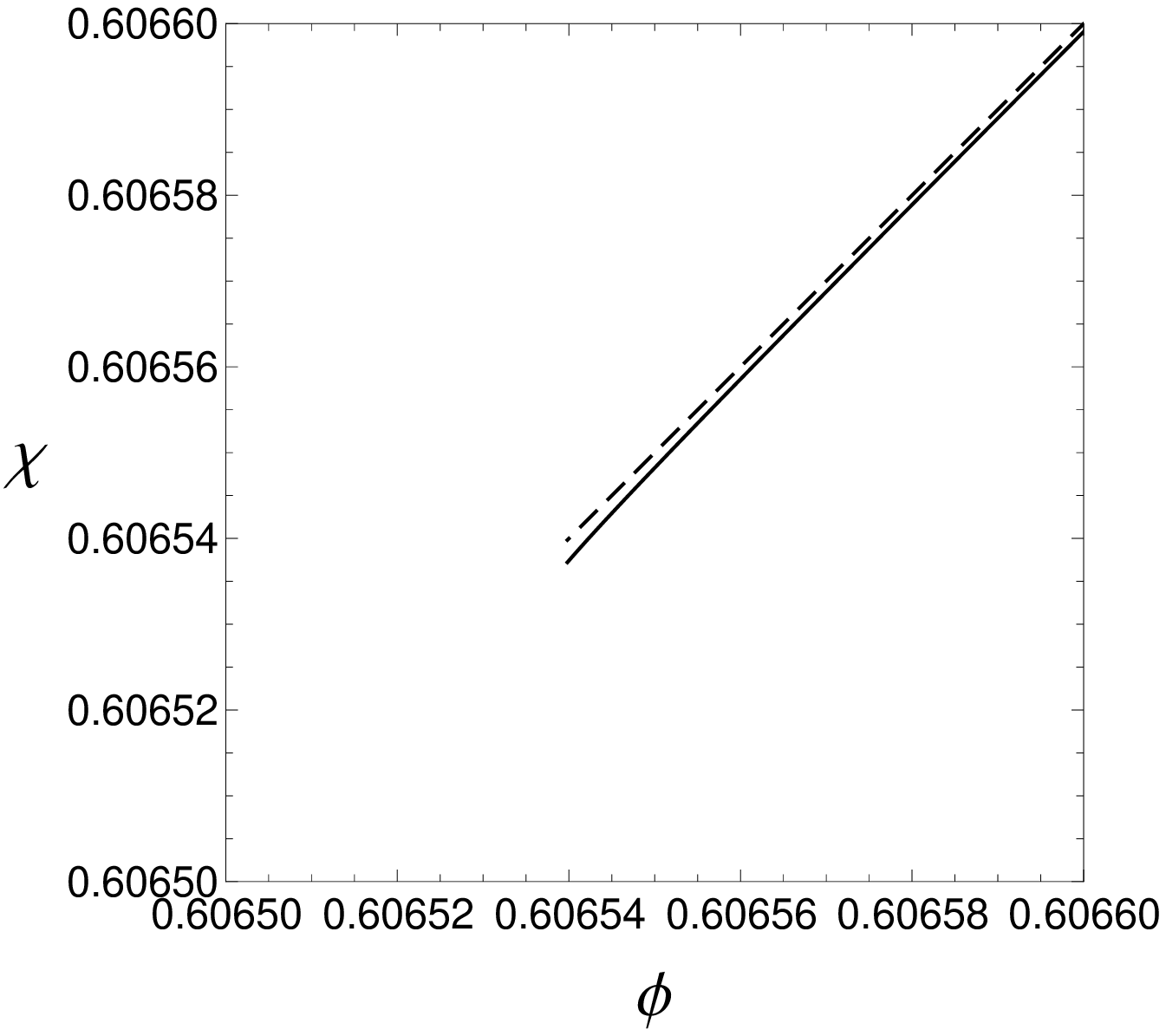}
      \caption{\emph{Left:} The trajectory in field space, originating at $\phi(t_{\star})=\chi(t_{\star})=1$ and ending after $N\simeq60$ e-folds of inflation, when $\epsilon=1$. \emph{Right:} Enlarged region of the trajectory (solid) illustrating the turn in the $\chi$ direction towards the end of inflation. The dashed line shows the straight line trajectory corresponding to $\mathcal{H}^{(\phi)}=\mathcal{H}^{(\chi)}$ for comparison.\label{fig:field-trajectory}}
      \end{center}
      \end{figure}

      \vspace{3pt}
      \be c^{(\phi)}=\frac{1}{\sqrt{1+\frac{4\mathcal{H}^{(\phi)^{2}}\lambda_{1}}{\phi^{4}}\left(1+\lambda_{2}\,\log\left(\frac{\phi}{\lambda_{3}}\right)\right)}},\ee
      \vspace{6pt}

\noindent where $c^{(\chi)}$ is again given by replacing $\phi\rightarrow\chi$. We then solve the background field equations (\ref{eq:field-dbi}) with $\lambda_{1}=6\times10^{16}$, which is typically required in standard DBI \cite{Alishahiha04}, $\lambda_{2}=2$ and $\lambda_{3}=1$. Furthermore, we choose $\phi(t_{\star})=\chi(t_{\star})=1$ as our initial conditions, where $t_{\star}$  is  the time at which observable scales exit the horizon. We require $N\simeq60$ e-folds between $t_{\star}$ and the end of inflation, which we choose to define as $\epsilon=1$. Finally we introduce a small asymmetry such that $\mathcal{H}^{(\phi)}=1.188\times10^{-6}$ and $\mathcal{H}^{(\chi)}=1.192\times10^{-6}$.  With this choice of parameters $c^{(\phi)}/c^{(\chi)}\sim1+10^{-3}$ at $t_{\star}$ and is therefore consistent with our approximation that the sound horizons are comparable when observable scales exit. Given this, we solve the field equations and plot the inflationary trajectory in figure~\ref{fig:field-trajectory}. Whilst approximately straight for most of inflation, the trajectory finally curves in the $\chi$ direction as the sound speeds rapidly increase in the tip regions of the throats. The turn in the trajectory is due to our choice of $\mathcal{H}^{(\chi)}>\mathcal{H}^{(\phi)}$, such that  $c^{(\chi)}$ increases slightly before $c^{(\phi)}$ towards the end of inflation.

\begin{figure}[t]
      \begin{center}
      \includegraphics[scale=0.43]{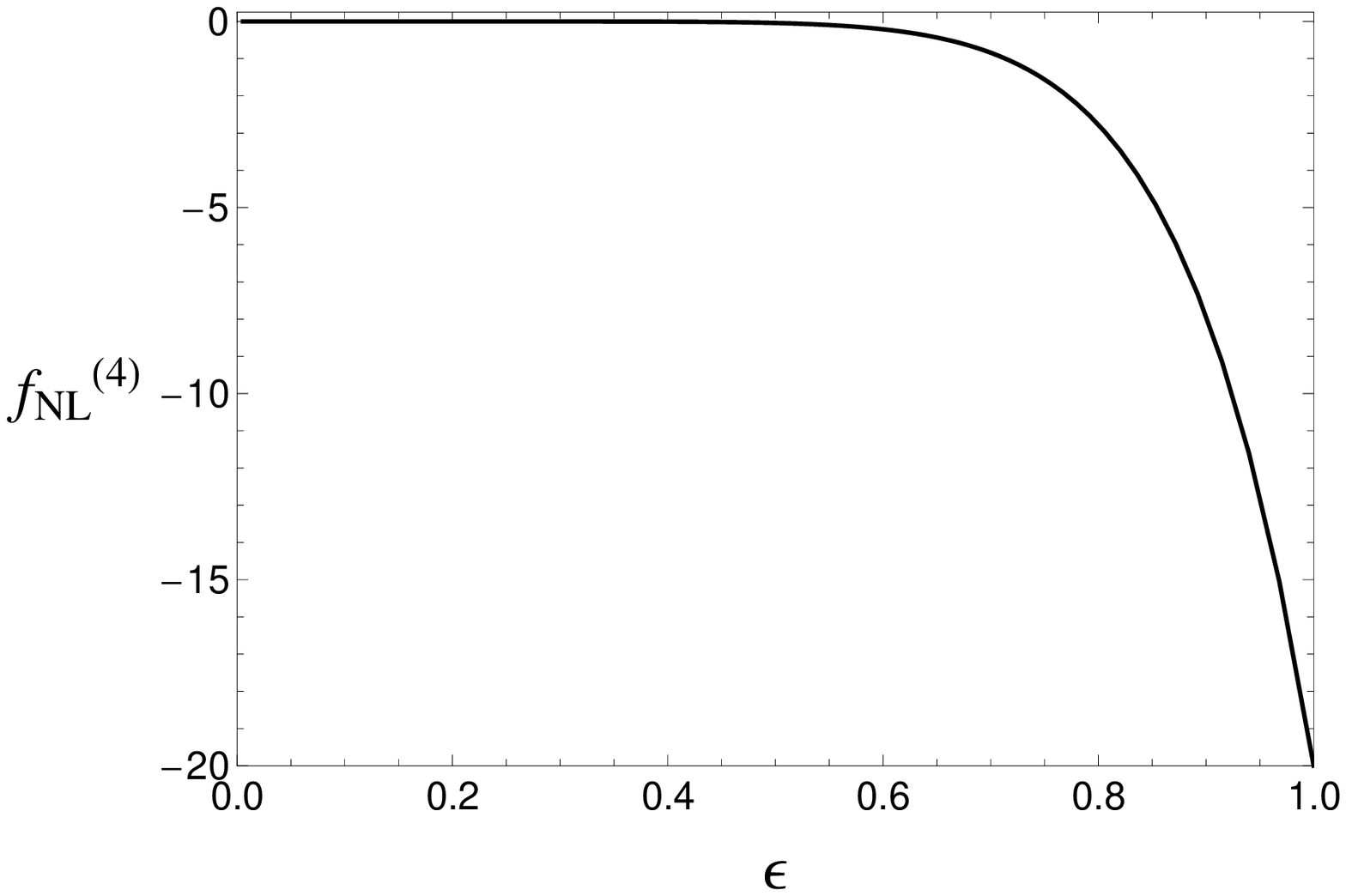}
      \includegraphics[scale=0.43]{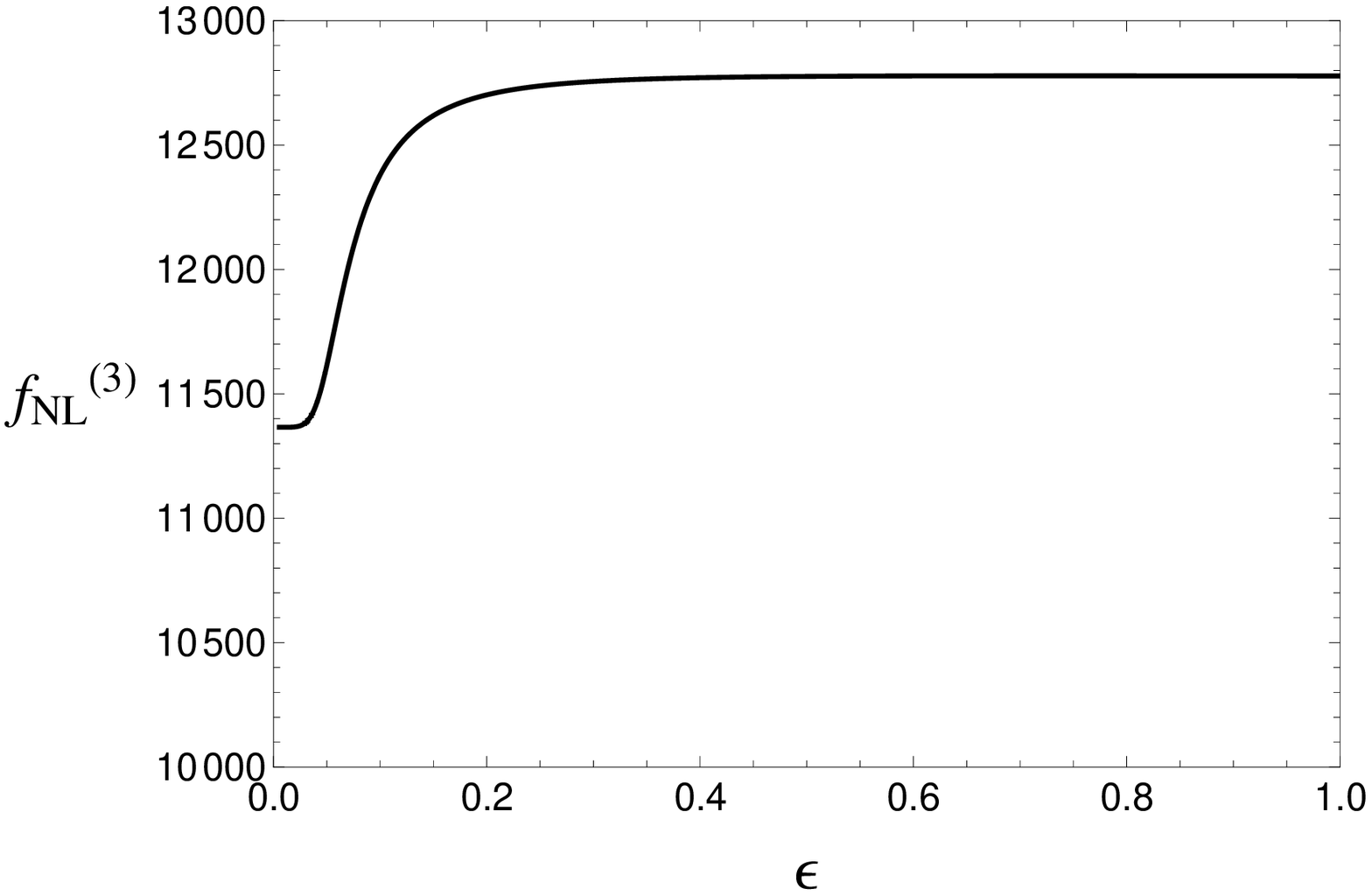}
      \caption{\emph{Left:} Evolution of $\fnl^{(3)}$ with respect to $t_{f}$, plotted as a function of $\epsilon$ for the trajectory in figure~\ref{fig:field-trajectory}. The background evolution after horizon exit enhances the initial value, producing $\fnl^{(3)}\simeq 12700$ at the end of inflation.  \emph{Right:} The evolution of $\fnl^{(4)}$, where the rapidly varying sound speeds produce $\fnl^{(4)}\simeq -20$ at the end of inflation.\label{fig:fnl}}
      \end{center}
      \end{figure}

\noindent Given this trajectory, we study the evolution of the relevant quantities as a function of $t_{f}$ for a fixed $t_{\star}$. We find that the quantities associated with the two-point statistics are consistent with observations \cite{Komatsu10}, where $\mathcal{P}_{\zeta}=2.44\times10^{-9}$ and $n_{\zeta}-1=-0.0108$ at the end of inflation (see \cite{Emery12} for the full expressions of these observables in the separable Hubble approach). With respect to the three-point function, figure \ref{fig:fnl} shows the evolution of $\fnl^{(4)}$ and $\fnl^{(3)}$ in the equilateral configuration, calculated using (\ref{eq:fnl4}) and (\ref{eq:fnl3}) respectively, as a function of the slow roll parameter $\epsilon$. As discussed in \cite{Emery12}, the rapidly increasing sound speeds in the tip of the throats produce $\fnl^{(4)}\simeq-20$ at the end of inflation. Looking now at the equilateral contribution, we see that the value of $c_{\star}\sim 10^{-2}$ provides $\fnl^{(3)}\sim 10^{4}$ at horizon exit. Thereafter however, the background evolution actually enhances this value between horizon exit and the end of inflation, eventually giving $\fnl^{(3)}\simeq 12700$. We conclude then that this scenario is inconsistent with current observations \cite{Komatsu10}. Whilst this need not necessarily be the case for all parameter values, we leave a fuller exploration of the parameter space to future work. 

\subsection{Exponential warp factors\label{sec:models-exponential}}

In this section we consider a phenomenologically similar, analytic model that is able to produce both equilateral and local non-Gaussianity that satisfies current constraints. Let us consider the following separable Hubble parameter
\be H=H_0\,\left(1- \frac{A_\phi}{2} e^{-\alpha \phi}- \frac{A_\chi}{2} e^{-\beta \chi} \right)\,. \ee
\noindent  We make the following choice for the warp factors
\bea
f^{(\phi)}&=&-\frac{e^{2\alpha\phi}}{\alpha^2\,A_\phi^2\,H_0^2}\,\left(1-\frac{e^{2( \gamma-1) \alpha \phi}}{B_\phi^2} \right),\\
f^{(\chi)}&=&-\frac{e^{2\beta\chi}}{\beta^2\,A_\chi^2\,H_0^2}\,\left(1-\frac{e^{2 (\delta-1) \beta \chi}}{B_\chi^2} \right)\,.
\eea
\noindent All the parameters in the previous formulae  are positive.
 This
choice of warp factors  is motivated by the fact that it will lead to an analytically solvable system.
The sound speeds read
\bea
c^{(\phi)}&=& B_\phi\,e^{-(\gamma-1) \alpha \phi},\\
c^{(\chi)}&=& B_\chi\,e^{-(\delta-1) \beta \chi}.
\eea
\noindent The equation of motion for the scalar $\phi$ results
\be
\dot{\phi}\,=\,-2 c^{(\phi)}\,H_{,\phi}\,=\,- H_0\,\alpha \,A_\phi\,B_\phi\,        e^{-\gamma\, \alpha \phi},
\ee
\noindent where the solution, assuming $\phi(0)=\phi_\star$, is
\bea\label{solphi}
\phi(t)&=&\phi_\star\,+\frac{1}{\alpha\,\gamma}\,\ln{\left[
1-\left(\alpha^2\, H_0\,\gamma \,A_\phi \,B_\phi\, e^{- \gamma\,\alpha\,\phi_\star}\right)\,t \right]}
\eea
\noindent while an analogous solution can be found  for $\chi(t)$. The solution for the scalar field is a decreasing function of time $t$. The  speeds of sound are increasing functions of time, provided
that $\gamma$ and $\delta$ are larger than one. The slow variation parameters read (in the limit in which ${A_\phi}\, e^{-\alpha \phi}\ll1$)
\bea
\epsilon^{(\phi)}&=&\frac{1}{2}\,\frac{H_0^2}{H^2}\, B_{\phi}\,\alpha^2\,A^2_{\phi}\,e^{-(\gamma+1)\alpha \phi},\\
\eta^{(\phi)}&=&\,-\frac{H_0}{H}\,  B_{\phi}\,\alpha^2\,A_{\phi}\,e^{- \gamma\,\alpha \phi},\\
s^{(\phi)}&=&\,- \frac{H_0}{H}\, B_{\phi}\,\left(\gamma-1\right)\,\alpha^2\,A_{\phi}\,e^{-\gamma\,\alpha \phi},\\
\eea
\noindent and we note the useful relations
\bea
\eta^{(\phi)}&=&\,-\sqrt{2 \,c^{(\phi)}}\,\alpha\,\sqrt{\epsilon^{(\phi)}},\\
s^{(\phi)}&=&\,-\sqrt{2 \,c^{(\phi)}}\,\left(\gamma-1\right)\,\alpha\,\sqrt{\epsilon^{(\phi)}},\\
H&=&\frac{H_0}{1-\frac{\eta^{(\phi)}}{\alpha^2 c^{(\phi)}   } -\frac{\eta^{(\chi)}}{\beta^2 c^{(\chi)} }}\,.
\eea

\noindent The number of e-folds is then given by
 \bea
 N_{e}&=&
 \frac{1}{2 \alpha^2\,\left(\gamma-1\right)}\left(\frac{1}{c_c^{(\phi)}}-\frac{1}{c_\star^{(\phi)}} \right)+
 \frac{1}{2 \beta^2\,\left(\delta-1\right)}\left(\frac{1}{c_c^{(\chi)}}-\frac{1}{c_\star^{(\chi)}} \right)
 \\
 &-&\frac{1}{\gamma}\left(\frac{1}{\eta^{(\phi)}_\star}-\frac{H_0}{H_c\,\eta^{(\phi)}_c} \right)
 -\frac{1}{\delta}\left(\frac{1}{\eta^{(\chi)}_\star}-\frac{H_0}{H_c\,\eta^{(\chi)}_c} \right). 
 \eea
 
We would like to find an inflationary trajectory for which the amplitude of the non-Gaussianity parameter $\fnl^{(3)}$ can be tuned sufficiently small to satisfy present constraints. At the same time, we would like that the amplitude of the local non-Gaussianity parameter $\fnl^{(4)}$ is sufficiently large to be detectable in the future, but still satisfying present-day constraints. Recall the expression for $\fnl^{(3)}$  (\ref{eq:fnl3-separable}) where
 
 \bea\label{expu}
 u&=&\frac12+\frac{1}{2 \epsilon}\,\left( \epsilon^{(\phi)} -  \epsilon^{(\chi)}+\frac{\epsilon^{(\chi)} \sqrt{\epsilon^{(\phi)}}}{\alpha \,\sqrt{c^{(\phi)}}}
 -\frac{\epsilon^{(\phi)} \sqrt{\epsilon^{(\chi)}}}{\beta \,\sqrt{c^{(\chi)}}}
 \right),
 \\
 v&=&1-u,
 \eea
 
\noindent in the approximation $\epsilon_\star\ll1$. If the speed of sound $c_\star\ll1$, the amplitude of  $\fnl^{(3)}$ is prohibitively large, unless $u$ is tuned in such a way that, at the end of inflation, the following inequality is satisfied
\\
\be\label{defsu}
\sigma(u)\,\equiv\,
\frac{\left(\frac{u^{3}}{\epsilon_{\star}^{(\phi)^{2}}}+\frac{\left(1-u\right)^{3}}{\epsilon_{\star}^{(\chi)^{2}}}\right)}{\left(\frac{u^{2}}{\epsilon_{\star}^{(\phi)}}+\frac{\left(1-u\right)^{2}}{\epsilon_{\star}^{(\chi)}}\right)^{2}}\,\ll\,1.
\ee

\noindent Recall also the expression for $\fnl^{(4)}$ (\ref{eq:fnl4-separable}), where the dominant contribution is given by   
\be
  \fnl^{(4)}\,=\,\frac{2\left( \frac{u}{\epsilon^{(\phi)_\star}} -  \frac{v}{\epsilon^{(\chi)_\star}}\right)^2}{\left( \frac{u^2}{\epsilon^{(\phi)}_\star} +
    \frac{v^2}{\epsilon^{(\chi)}_\star} \right)^2}{\cal A},
  \ee
\noindent where
\bea
  {\cal A}&=&\frac{H_f^2}{H_\star^2}\,\frac{\epsilon_f^{(\phi)} \epsilon_f^{(\chi)} }{\epsilon_f}
  \,\left(\frac{\eta_f^{ss}}{\epsilon_f}+ \frac{s_f^{ss}}{2\,\epsilon_f} \right).
  \eea
\noindent In order to obtain a detectable $\fnl^{(4)}$, we have to find situations in which either or both 
$ \eta^{ss}$ and $s^{ss}$ are large at the end of inflation.

Let us start discussing the conditions for being able to tune the value of $\fnl^{(3)}$. We assume the following hierarchy between the slow-roll parameters at horizon exit

\be\epsilon_{\star}^{(\phi)}\,\equiv \,r\,\epsilon_{\star}^{(\chi)},\hskip1cm {\text{with\,\,}}r\, >\,1.\label{defr} \ee
 
\noindent The function $\sigma(u)$, defined in (\ref{defsu}), vanishes at the point ${r^\frac23}/{(r^\frac23-1)} $. We then demand that the parameter $u$ satisfies
\be\label{defue}u_{end}\,=\,\frac{r^\frac23}{r^\frac23-1}+\lambda ,\ee
\noindent at the end of inflation, where $\lambda$ small in absolute value. Expanding the function $\sigma(u)$  at first order in $\lambda$, one finds 

 \be \sigma(u_{end})\,=\,-\frac{(r^{1/3}+1) (r^{1/3}-1)^3}{r^{2/3}}\,\lambda .\ee
 \noindent We can therefore set  $\sigma(u_{end})$ to be sufficiently small to compensate the enhancement associated with the speed of sound in (\ref{eq:fnl3-separable}).

 Let us assume that, at the end of inflation, $\epsilon^{(\phi)}$ is much larger than $ \epsilon^{(\chi)}$

 \be\label{relee}\epsilon^{(\chi)}\,=\,\ka_1\,\epsilon^{(\phi)} \hskip0.5cm,\hskip0.5cm {\text{ with $\ka_1\ll1$.}}\ee

\noindent Moreover, we write the following expressions for the speeds of sound at the end of inflation
 \be
 c^{(\phi)}\,=\,\frac{\ka_2^2}{\alpha^2}\,\epsilon^{(\phi)},\hskip1cm
 c^{(\chi)}\,=\,\frac{\ka_3^2}{\beta^2}\,\epsilon^{(\phi)},\label{css}
 \ee
\noindent such that we can write

 \be
 u_{end}=1-\frac{\ka_1}{1+\ka_1}+\left(\frac{\sqrt{\ka_1}}{\ka_2}-\frac{1}{\ka_3}\right)\,\frac{\sqrt{\ka_1}}{1+\ka_1}
 \ee
 
\noindent  Tuning properly the $\ka_i$  then, this quantity can assume the desired value of (\ref{defue}). 
 
We now consider  possible inflationary trajectories, with the specific requirements listed above, that lead to observationally viable non-Gaussianities of both local and equilateral type. We would like to determine which conditions we have to impose to the model parameters in order to satisfy all our requirements. For simplicity,  we assume that the $\gamma$ and $\delta$ are very large, so that we can safely neglect corrections weighted by inverse powers of these parameters. We proceed by discussing one by one  the conditions that fix our  parameters. As discussed in the previous sections,  in all our analysis we make the hypothesis that, at horizon exit, $c_\star^{(\phi)}\,=\,c_\star^{(\chi)}\,=\,c_\star$. This implies 

\be
\phi_\star\,=\, \frac{{\left(\delta-1\right)}}{{\left(\gamma-1\right)}}\,\frac{\beta}{\alpha}\,\chi_\star
+\frac{1}{\alpha\,(\gamma-1)}\,\ln{\left(\frac{B_\phi}{B_\chi}\right)}.
\ee

\noindent The quantity $\phi_\star$ is associated with the speed of sound $c_\star$ at horizon exit by

\be
\phi_\star\,=\,\frac{1}{\alpha\,\left(\gamma-1\right)}\,\ln{\left(\frac{B_\phi}{c_\star}\right)}.
\ee

\noindent The condition (\ref{defr}) gives
\be\label{conA}
A_\phi\,=\,\sqrt{r}\,\frac{\beta}{\alpha}\,A_\chi ,
\ee
\noindent where we neglect corrections that scale as $1/\gamma$ and $1/\delta$ since, as we assumed above, these are negligible. Condition (\ref{css}) gives
   
\bea
A_\phi&=&\frac{\sqrt{2}}{\ka_2}\frac{1}{\left( 1+\frac{\sqrt{2}}{\alpha \,\ka_2}+\frac{\sqrt{2 \ka_1}}{\beta\,\ka_3}\right)},
\\
A_\chi&=&\frac{\sqrt{2}}{\ka_3}\frac{1}{\left( 1+\frac{\sqrt{2}}{\alpha \,\ka_2}+\frac{\sqrt{2 \ka_1}}{\beta\,\ka_3}\right)}.
\eea

\noindent Hence, combining with (\ref{conA}), we find

\be\label{ccond}
 \frac{\alpha}{\beta}\,=\,\sqrt{r}\,\frac{\ka_2}{\ka_3}\,.
 \ee

\noindent In the approximation $H\simeq H_0$,  the slow-roll parameters at horizon exit are given by 

\bea
\epsilon_\star^{(\phi)}&\simeq& \frac{\alpha^2}{2 \,k_2^2}\,c_\star , \hskip1cm \epsilon_\star^{(\chi)}\,=\,\frac{1}{r}\, \epsilon_\star^{(\phi)},\\
\eta_\star^{(\phi)}&\simeq&- \frac{\alpha^2}{k_2}\,c_\star ,\hskip1cm \eta_\star^{(\chi)}\,\simeq\,- \frac{\beta^2}{k_3}\,c_\star ,\\
s_\star^{(\phi)}&\simeq& \frac{\alpha^2\,{\gamma}}{k_2 }\,c_\star ,\hskip1cm s_\star^{(\chi)}\,
\simeq\, \frac{\beta^2\,{\delta}}{k_3}\,c_\star .
\eea

\noindent The dominant contributions to the  number of e-folds depends on the terms evaluated at horizon exit and with the aid of the previous equations we find
\be
N_e\,\simeq\,
\frac{1}{2}\,\left( \frac{1}{s_\star^{(\phi)}}+\frac{1}{s_\star^{(\chi)}}\right).
\ee

\noindent
The dominant contributions to the non-Gaussian parameters in this scenario read as follows (we write only their amplitude, and not the scale dependence)
\bea
\fnl^{(3)}&=&-\frac{5}{6}\,\frac{1}{c_\star^2}\,\frac{\left(r^{1/3}+1\right)\left(r^{1/3}-1\right)^3}{r^{2/3}}\,\lambda\label{ref3}\,,\\
\fnl^{(4)}&=&-\frac{2\left(r^{2/3}-1\right)^2}{r^{2/3}}\,\frac{\ka_1\,\left(\delta \beta+\ka_1 \gamma\alpha \right)}{\left( 1+\frac{\sqrt{2}}{\alpha \,\ka_2}+\frac{\sqrt{2 \ka_1}}{\beta\,\ka_3}\right)^2}\label{ref4}\,.
\eea

\bigskip

\noindent Without providing an exhaustive analysis, let us consider a concrete set-up in which we assign the following numerical values to the parameters $\ka_i$
\bea
\ka_1&=&10^{-3}\,,\\
\ka_2&=&10^{-1}\,,\\
\ka_3&=&10\,.
\eea

\noindent We demand also that 
\be s_\star^{(\phi)}\,\simeq s_\star^{(\chi)}\,\simeq\,10^{-2},\label{conos}\ee 

\noindent to obtain a sufficient number of e-folds. With this choice, the value of $u_{end}$ is 
$$
u_{end}\,\simeq\,1+6\times 10^{-3}\,,
$$
\noindent implying that (see (\ref{defue}))  $$r\simeq 2.3\times10^3+5.8\times10^5\,\lambda\,.$$ Hence,
assuming that $\lambda < 10^{-3}$ (we will see that this assumption is satisfied in our set-up) 
 (\ref{ccond}) gives
\be
\beta\simeq 2\,\alpha\,.
\ee
  
\noindent The condition (\ref{conos}) then provides the relations
  
\be\label{inci}
  c_\star\simeq \frac{10^{-2}\,\ka_2}{\alpha^2\,\gamma}\,\simeq\, \frac{10^{-2}\,\ka_3}{\beta^2\,\delta}, 
  \ee

\noindent implying  that $\delta\simeq 10\gamma$. Let us consider the local non-Gaussian parameter $\fnl^{(4)}$:
using (\ref{ref4}) we find
\bea
\fnl^{(4)}&\simeq&-
3\times 10^{-5}\,\alpha^3\,\delta\,.
\eea
Since our parameters $\alpha$ and $\delta$ are positive, this quantity is always negative. Choosing for definiteness
 \be\delta\,\simeq\,\frac{10^5}{3\,\alpha^3}\label{delal}\ee
we get
$$|\fnl^{(4)}|\,=\,{\cal O}(1)\,.$$

\noindent
Substituting the information of (\ref{delal})
into (\ref{inci}), we get 
\be
c_\star\,=\,3\,\alpha\,\times10^{-7}\,.
\ee

\noindent
We are not allowed to choose
  too large values for $\alpha$, since condition (\ref{delal}) would lead to small values of $\delta$, against our
working hypothesis. We set $\alpha=3$, that gives
$$
 c_\star\,=\,10^{-6}\,.
$$
Hence, $\fnl^{(3)}$ is easily obtained from (\ref{ref3}) by substituting our values of the parameters:
\be
\fnl^{(3)}\,\simeq\,-1.4\,\lambda\times 10^8\,.
\ee

 By tuning appropriately the parameter $\lambda$ to a value of order  $10^{-7}$ one finds
 
\be| \fnl^{(3)}|\,\simeq\,{\cal O}(10).\ee

\noindent
and its sign depends on the sign of $\lambda$. 
  Hence, this  semi-quantitative analysis shows that within this concrete model we can obtain relatively large non-Gaussianities both of local and equilateral type. It would be interesting to perform a more complete analysis of this set-up to determine more precisely its predictions. Moreover, since the exponential warp factors were introduced for analytical ease, it would also be prudent to use these results as a starting point for further parameter exploration of the previous example (section \ref{sec:models-cutoff}), which whilst more realistic is less analytically tractable.   

\section{Conclusions\label{sec:conclusions}}

We analysed a setup of multiple-field DBI inflation leading to mixed form
of primordial non-Gaussianity, including equilateral and local
bispectrum shapes.
Previously, we studied a multiple-DBI model as an example of multi-component inflation with non-standard kinetic terms, showing that rapidly varying sound speeds can produce large local type non-Gaussianity during a turn in the
 trajectory \cite{Emery12}. Here we have included the equilateral contribution produced on sub-horizon scales, and found the possibility of an observationally viable mixed non-Gaussian signal. 

 We used a general formalism based on the
Hamilton-Jacobi approach, allowing us to go beyond slow-roll,
combining the three-point function for the fields at Hubble-exit  with
the non-linear evolution of super-Hubble scales calculated using
the $\delta N$ formalism.  
We were able to obtain
 analytic results by taking a separable Ansatz for the Hubble rate. We
found general expressions for both the equilateral and local type
non-Gaussianity parameter $\fnl$. The equilateral non-Gaussianity
includes the usual enhancement for small sound speeds, but multiplied
by an analytic factor which can lead to a suppression of this quantity. 
This modulation may suppress the value of $\fnl^{(3)}$ to within observational bounds
 (see \cite{Langlois08b, Kidani12} for analogous results).

 We applied
our findings in two explicit scenarios. In the first model, previously found
to have detectable local non-Gaussianity, we found that the equilateral
signal is not sufficiently suppressed to evade current observational
bounds. In our second set-up we constructed an observationally viable model which exhibits both
  equilateral $\fnl^{(3)}$ and a negative local $\fnl^{(4)}$,
  providing a first step towards understanding regimes in which this can occur in more realistic scenarios.

Open issues remain however, the most prevalent being a better understanding of the likelihood of such a mixed non-Gaussian signal. For example, it would be interesting to use the results of our analytical example in section~\ref{sec:models-exponential} to more methodically assess the parameter space of the phenomenologically similar but more realistic scenario in section~\ref{sec:models-cutoff}. Moreover, a consistency relation between the more general formulae for $\fnl^{(3)}$ (\ref{eq:fnl3-separable}) and $\fnl^{(4)}$ (\ref{eq:fnl4-separable}) would be desirable, since this would be more widely applicable. The use of a sum separable Hubble parameter, whilst providing analytically tractable expressions, also restricts the choice of potential. Thus it would be  more general still to consider alternative analytical or numerical methods. Finally, we again note that our approach only treats multiple-field dynamics during inflation and we cannot necessarily conclude that such values are observed in the CMB. It would be prudent therefore to use analogous techniques to \cite{Elliston11a,Elliston11b,Meyers11-a,Meyers11-b,Peterson11a,Watanabe11,Choi12,Mazumdar12} to further our understanding of inflationary dynamics from observations of non-Gaussianity in the CMB. 

We conclude by noting two further directions to develop the above, in addition to increased generality. In our previous work we made a first attempt at addressing the next order statistic, the trispectrum, by plotting the evolution of the analogues of the non-linearity parameter, $g_{\textrm{NL}}$ and $\tau_{\textrm{NL}}$, in the example of inflation in two cut-off throats. In light of impending observations by Planck \cite{Planck06} however, it would be interesting to further characterise the model by more rigorously considering its predictions for the trispectrum. Finally, we note again that in the third order action (\ref{eq:action-third-leading-DBI}) we worked to leading order in small sound speeds at horizon exit. It is conceivable however that the next to leading order terms may still provide a significant contribution, with the potential to generate shapes distinct from the local and equilateral types. Since tentative signals for the orthogonal shape have been detected by WMAP \cite{Bennett12}, and again given the impending results from Planck, it would be interesting to consider such contributions in future work.

\acknowledgments

The authors would like to thank Taichi Kidani and Guido W. Pettinari for useful discussions. JE is supported by an STFC doctoral training grant ST/F007531/1. GT is supported by an STFC Advanced Fellowship ST/H005498/1. DW is supported by STFC grant ST/H002774/1.


\begin{thebibliography}{10}

\bibitem{Lyth99}
D.~H. Lyth and A.~Riotto, {\it {Particle physics models of inflation and the
  cosmological density perturbation}},  {\em Phys. Rept.} {\bf 314} (1999)
  1--146, [\href{http://xxx.lanl.gov/abs/hep-ph/9807278}{{\tt
  hep-ph/9807278}}].

\bibitem{Lyth09}
D.~H. Lyth and A.~Liddle, {\em The primordial density perturbation: cosmology,
  inflation and the origin of structure}.
\newblock Cambridge Univ. Press, Cambridge, 2009.

\bibitem{Mazumdar10}
A.~Mazumdar and J.~Rocher, {\it {Particle physics models of inflation and
  curvaton scenarios}},  {\em Phys. Rept.} {\bf 497} (2011) 85--215,
  [\href{http://xxx.lanl.gov/abs/1001.0993}{{\tt arXiv:1001.0993}}].

\bibitem{Komatsu09}
E.~Komatsu, N.~Afshordi, N.~Bartolo, D.~Baumann, J.~Bond, {\em et.~al.}, {\it
  {Non-Gaussianity as a Probe of the Physics of the Primordial Universe and the
  Astrophysics of the Low Redshift Universe}},
  \href{http://xxx.lanl.gov/abs/0902.4759}{{\tt arXiv:0902.4759}}.

\bibitem{Planck06}
{The Planck Collaboration}, {\it {The Scientific programme of planck}},
  \href{http://xxx.lanl.gov/abs/astro-ph/0604069}{{\tt astro-ph/0604069}}. Also
  available for direct download from http://www.rssd.esa.int/Planck Report-no:
  ESA-SCI(2005)1.

\bibitem{Bartolo04}
N.~Bartolo, E.~Komatsu, S.~Matarrese, and A.~Riotto, {\it {Non-Gaussianity from
  inflation: Theory and observations}},  {\em Phys. Rept.} {\bf 402} (2004)
  103--266, [\href{http://xxx.lanl.gov/abs/astro-ph/0406398}{{\tt
  astro-ph/0406398}}].

\bibitem{Chen10}
X.~Chen, {\it {Primordial Non-Gaussianities from Inflation Models}},  {\em Adv.
  Astron.} {\bf 2010} (2010) 638979,
  [\href{http://xxx.lanl.gov/abs/1002.1416}{{\tt arXiv:1002.1416}}].

\bibitem{Moroi01}
T.~Moroi and T.~Takahashi, {\it {Effects of cosmological moduli fields on
  cosmic microwave background}},  {\em Phys. Lett.} {\bf B 522} (2001)
  215--221, [\href{http://xxx.lanl.gov/abs/hep-ph/0110096}{{\tt
  hep-ph/0110096}}].

\bibitem{Lyth02}
D.~H. Lyth and D.~Wands, {\it {Generating the curvature perturbation without an
  inflaton}},  {\em Phys. Lett.} {\bf B 524} (2002) 5--14,
  [\href{http://xxx.lanl.gov/abs/hep-ph/0110002}{{\tt hep-ph/0110002}}].

\bibitem{Lyth03}
D.~H. Lyth and D.~Wands, {\it {Conserved cosmological perturbations}},  {\em
  Phys. Rev.} {\bf D 68} (2003) 103515,
  [\href{http://xxx.lanl.gov/abs/astro-ph/0306498}{{\tt astro-ph/0306498}}].

\bibitem{Kofman03}
L.~Kofman, {\it {Probing string theory with modulated cosmological
  fluctuations}},  \href{http://xxx.lanl.gov/abs/astro-ph/0303614}{{\tt
  astro-ph/0303614}}.

\bibitem{Dvali04}
G.~Dvali, A.~Gruzinov, and M.~Zaldarriaga, {\it {A new mechanism for generating
  density perturbations from inflation}},  {\em Phys. Rev.} {\bf D 69} (2004)
  023505, [\href{http://xxx.lanl.gov/abs/astro-ph/0303591}{{\tt
  astro-ph/0303591}}].

\bibitem{Burgess10}
C.~Burgess, M.~Cicoli, M.~Gomez-Reino, F.~Quevedo, G.~Tasinato, {\em et.~al.},
  {\it {Non-standard primordial fluctuations and nongaussianity in string
  inflation}},  {\em JHEP} {\bf 1008} (2010) 045,
  [\href{http://xxx.lanl.gov/abs/1005.4840}{{\tt arXiv:1005.4840}}].

\bibitem{Cicoli12}
M.~Cicoli, G.~Tasinato, I.~Zavala, C.~Burgess, and F.~Quevedo, {\it {Modulated
  Reheating and Large Non-Gaussianity in String Cosmology}},  {\em JCAP} {\bf
  1205} (2012) 039, [\href{http://xxx.lanl.gov/abs/1202.4580}{{\tt
  arXiv:1202.4580}}].

\bibitem{Gordon01}
C.~Gordon, D.~Wands, B.~A. Bassett, and R.~Maartens, {\it {Adiabatic and
  entropy perturbations from inflation}},  {\em Phys. Rev.} {\bf D 63} (2001)
  023506, [\href{http://xxx.lanl.gov/abs/astro-ph/0009131}{{\tt
  astro-ph/0009131}}].

\bibitem{Nibbelink02}
S.~{Groot Nibbelink} and B.~{van Tent}, {\it {Scalar perturbations during
  multiple field slow-roll inflation}},  {\em Class. Quant. Grav.} {\bf 19}
  (2002) 613--640, [\href{http://xxx.lanl.gov/abs/hep-ph/0107272}{{\tt
  hep-ph/0107272}}].

\bibitem{Rigopoulos04}
G.~Rigopoulos, {\it {On second order gauge invariant perturbations in
  multi-field inflationary models}},  {\em Class. Quant. Grav.} {\bf 21} (2004)
  1737--1754, [\href{http://xxx.lanl.gov/abs/astro-ph/0212141}{{\tt
  astro-ph/0212141}}].

\bibitem{Wands00}
D.~Wands, K.~A. Malik, D.~H. Lyth, and A.~R. Liddle, {\it {A New approach to
  the evolution of cosmological perturbations on large scales}},  {\em Phys.
  Rev.} {\bf D 62} (2000) 043527,
  [\href{http://xxx.lanl.gov/abs/astro-ph/0003278}{{\tt astro-ph/0003278}}].

\bibitem{Lyth05}
D.~H. Lyth, K.~A. Malik, and M.~Sasaki, {\it {A General proof of the
  conservation of the curvature perturbation}},  {\em JCAP} {\bf 0505} (2005)
  004, [\href{http://xxx.lanl.gov/abs/astro-ph/0411220}{{\tt
  astro-ph/0411220}}].

\bibitem{Bernardeau02}
F.~Bernardeau and J.-P. Uzan, {\it {Non-Gaussianity in multifield inflation}},
  {\em Phys. Rev.} {\bf D 66} (2002) 103506,
  [\href{http://xxx.lanl.gov/abs/hep-ph/0207295}{{\tt hep-ph/0207295}}].

\bibitem{Vernizzi06}
F.~Vernizzi and D.~Wands, {\it {Non-gaussianities in two-field inflation}},
  {\em JCAP} {\bf 0605} (2006) 019,
  [\href{http://xxx.lanl.gov/abs/astro-ph/0603799}{{\tt astro-ph/0603799}}].

\bibitem{Rigopoulos06}
G.~Rigopoulos, E.~Shellard, and B.~van Tent, {\it {Large non-Gaussianity in
  multiple-field inflation}},  {\em Phys. Rev.} {\bf D 73} (2006) 083522,
  [\href{http://xxx.lanl.gov/abs/astro-ph/0506704}{{\tt astro-ph/0506704}}].

\bibitem{Rigopoulos07}
G.~Rigopoulos, E.~Shellard, and B.~{van Tent}, {\it {Quantitative bispectra
  from multifield inflation}},  {\em Phys. Rev.} {\bf D 76} (2007) 083512,
  [\href{http://xxx.lanl.gov/abs/astro-ph/0511041}{{\tt astro-ph/0511041}}].

\bibitem{Yokoyama07}
S.~Yokoyama, T.~Suyama, and T.~Tanaka, {\it {Primordial Non-Gaussianity in
  Multi-Scalar Slow-Roll Inflation}},  {\em JCAP} {\bf 0707} (2007) 013,
  [\href{http://xxx.lanl.gov/abs/0705.3178}{{\tt arXiv:0705.3178}}].

\bibitem{Battefeld07}
T.~Battefeld and R.~Easther, {\it {Non-Gaussianities in Multi-field
  Inflation}},  {\em JCAP} {\bf 0703} (2007) 020,
  [\href{http://xxx.lanl.gov/abs/astro-ph/0610296}{{\tt astro-ph/0610296}}].

\bibitem{Yokoyama08}
S.~Yokoyama, T.~Suyama, and T.~Tanaka, {\it {Primordial Non-Gaussianity in
  Multi-Scalar Inflation}},  {\em Phys. Rev.} {\bf D 77} (2008) 083511,
  [\href{http://xxx.lanl.gov/abs/0711.2920}{{\tt arXiv:0711.2920}}].

\bibitem{Byrnes08}
C.~T. Byrnes, K.-Y. Choi, and L.~M. Hall, {\it {Conditions for large
  non-Gaussianity in two-field slow-roll inflation}},  {\em JCAP} {\bf 0810}
  (2008) 008, [\href{http://xxx.lanl.gov/abs/0807.1101}{{\tt
  arXiv:0807.1101}}].

\bibitem{Sasaki08}
M.~Sasaki, {\it {Multi-brid inflation and non-Gaussianity}},  {\em Prog. Theor.
  Phys.} {\bf 120} (2008) 159--174,
  [\href{http://xxx.lanl.gov/abs/0805.0974}{{\tt arXiv:0805.0974}}].

\bibitem{Byrnes09}
C.~T. Byrnes and G.~Tasinato, {\it {Non-Gaussianity beyond slow roll in
  multi-field inflation}},  {\em JCAP} {\bf 0908} (2009) 016,
  [\href{http://xxx.lanl.gov/abs/0906.0767}{{\tt arXiv:0906.0767}}].

\bibitem{Battefeld09}
D.~Battefeld and T.~Battefeld, {\it {On Non-Gaussianities in Multi-Field
  Inflation (N fields): Bi and Tri-spectra beyond Slow-Roll}},  {\em JCAP} {\bf
  0911} (2009) 010, [\href{http://xxx.lanl.gov/abs/0908.4269}{{\tt
  arXiv:0908.4269}}].

\bibitem{Wang10}
T.~Wang, {\it {Note on Non-Gaussianities in Two-field Inflation}},  {\em Phys.
  Rev.} {\bf D 82} (2010) 123515,
  [\href{http://xxx.lanl.gov/abs/1008.3198}{{\tt arXiv:1008.3198}}].

\bibitem{Tzavara11}
E.~Tzavara and B.~van Tent, {\it {Bispectra from two-field inflation using the
  long-wavelength formalism}},  {\em JCAP} {\bf 1106} (2011) 026,
  [\href{http://xxx.lanl.gov/abs/1012.6027}{{\tt arXiv:1012.6027}}].

\bibitem{Elliston11a}
J.~Elliston, D.~Mulryne, D.~Seery, and R.~Tavakol, {\it {Evolution of
  non-Gaussianity in multi-scalar field models}},  {\em Int. J. Mod. Phys.}
  {\bf A 26} (2011) 3821--3832, [\href{http://xxx.lanl.gov/abs/1107.2270}{{\tt
  arXiv:1107.2270}}].

\bibitem{Elliston11b}
J.~Elliston, D.~J. Mulryne, D.~Seery, and R.~Tavakol, {\it {Evolution of fNL to
  the adiabatic limit}},  {\em JCAP} {\bf 1111} (2011) 005,
  [\href{http://xxx.lanl.gov/abs/1106.2153}{{\tt arXiv:1106.2153}}].

\bibitem{Meyers11-a}
J.~Meyers and N.~Sivanandam, {\it {Non-Gaussianities in Multifield Inflation:
  Superhorizon Evolution, Adiabaticity, and the Fate of fnl}},  {\em Phys.
  Rev.} {\bf D 83} (2011) 103517,
  [\href{http://xxx.lanl.gov/abs/1011.4934}{{\tt arXiv:1011.4934}}].

\bibitem{Meyers11-b}
J.~Meyers and N.~Sivanandam, {\it {Adiabaticity and the Fate of
  Non-Gaussianities: The Trispectrum and Beyond}},  {\em Phys. Rev.} {\bf D 84}
  (2011) 063522, [\href{http://xxx.lanl.gov/abs/1104.5238}{{\tt
  arXiv:1104.5238}}].

\bibitem{Peterson11a}
C.~M. Peterson and M.~Tegmark, {\it {Non-Gaussianity in Two-Field Inflation}},
  {\em Phys. Rev.} {\bf D 84} (2011) 023520,
  [\href{http://xxx.lanl.gov/abs/1011.6675}{{\tt arXiv:1011.6675}}].

\bibitem{Watanabe11}
Y.~Watanabe, {\it {$\delta N$ vs. covariant perturbative approach to non-
  Gaussianity outside the horizon in multi-field inflation}},
  \href{http://xxx.lanl.gov/abs/1110.2462}{{\tt arXiv:1110.2462}}.

\bibitem{Choi12}
K.-Y. Choi, S.~A. Kim, and B.~Kyae, {\it {Primordial curvature perturbation
  during and at the end of multi-field inflation}},
  \href{http://xxx.lanl.gov/abs/1202.0089}{{\tt arXiv:1202.0089}}.

\bibitem{Mazumdar12}
A.~Mazumdar and L.~Wang, {\it {Separable and non-separable multi-field
  inflation and large non-Gaussianity}},
  \href{http://xxx.lanl.gov/abs/1203.3558}{{\tt arXiv:1203.3558}}.

\bibitem{Frazer11a}
J.~Frazer and A.~R. Liddle, {\it {Multi-field inflation with random potentials:
  field dimension, feature scale and non-Gaussianity}},  {\em JCAP} {\bf 1202}
  (2012) 039, [\href{http://xxx.lanl.gov/abs/1111.6646}{{\tt
  arXiv:1111.6646}}].

\bibitem{Battefeld12}
D.~Battefeld, T.~Battefeld, and S.~Schulz, {\it {On the Unlikeliness of
  Multi-Field Inflation: Bounded Random Potentials and our Vacuum}},
  \href{http://xxx.lanl.gov/abs/1203.3941}{{\tt arXiv:1203.3941}}.

\bibitem{Chen07}
X.~Chen, M.~xin Huang, S.~Kachru, and G.~Shiu, {\it {Observational signatures
  and non-Gaussianities of general single field inflation}},  {\em JCAP} {\bf
  0701} (2007) 002, [\href{http://xxx.lanl.gov/abs/hep-th/0605045}{{\tt
  hep-th/0605045}}].

\bibitem{Koyama10}
K.~Koyama, {\it {Non-Gaussianity of quantum fields during inflation}},  {\em
  Class. Quant. Grav.} {\bf 27} (2010) 124001,
  [\href{http://xxx.lanl.gov/abs/1002.0600}{{\tt arXiv:1002.0600}}].

\bibitem{Christopherson09}
A.~J. Christopherson and K.~A. Malik, {\it {The non-adiabatic pressure in
  general scalar field systems}},  {\em Phys. Lett.} {\bf B 675} (2009)
  159--163, [\href{http://xxx.lanl.gov/abs/0809.3518}{{\tt arXiv:0809.3518}}].

\bibitem{Silverstein04}
E.~Silverstein and D.~Tong, {\it {Scalar speed limits and cosmology:
  Acceleration from D-cceleration}},  {\em Phys. Rev.} {\bf D 70} (2004)
  103505, [\href{http://xxx.lanl.gov/abs/hep-th/0310221}{{\tt
  hep-th/0310221}}].

\bibitem{Alishahiha04}
M.~Alishahiha, E.~Silverstein, and D.~Tong, {\it {DBI in the sky}},  {\em Phys.
  Rev.} {\bf D 70} (2004) 123505,
  [\href{http://xxx.lanl.gov/abs/hep-th/0404084}{{\tt hep-th/0404084}}].

\bibitem{Langlois08}
D.~Langlois, S.~Renaux-Petel, D.~A. Steer, and T.~Tanaka, {\it {Primordial
  perturbations and non-Gaussianities in DBI and general multi-field
  inflation}},  {\em Phys. Rev.} {\bf D 78} (2008) 063523,
  [\href{http://xxx.lanl.gov/abs/0806.0336}{{\tt arXiv:0806.0336}}].

\bibitem{Langlois08b}
D.~Langlois, S.~Renaux-Petel, D.~A. Steer, and T.~Tanaka, {\it {Primordial
  fluctuations and non-Gaussianities in multi-field DBI inflation}},  {\em
  Phys. Rev. Lett.} {\bf 101} (2008) 061301,
  [\href{http://xxx.lanl.gov/abs/0804.3139}{{\tt arXiv:0804.3139}}].

\bibitem{Arroja08}
F.~Arroja, S.~Mizuno, and K.~Koyama, {\it {Non-gaussianity from the bispectrum
  in general multiple field inflation}},  {\em JCAP} {\bf 0808} (2008) 015,
  [\href{http://xxx.lanl.gov/abs/0806.0619}{{\tt arXiv:0806.0619}}].

\bibitem{RenauxPetel09a}
S.~Renaux-Petel and G.~Tasinato, {\it {Nonlinear perturbations of cosmological
  scalar fields with non-standard kinetic terms}},  {\em JCAP} {\bf 0901}
  (2009) 012, [\href{http://xxx.lanl.gov/abs/0810.2405}{{\tt
  arXiv:0810.2405}}].

\bibitem{Gao09}
X.~Gao and B.~Hu, {\it {Primordial Trispectrum from Entropy Perturbations in
  Multifield DBI Model}},  {\em JCAP} {\bf 0908} (2009) 012,
  [\href{http://xxx.lanl.gov/abs/0903.1920}{{\tt arXiv:0903.1920}}].

\bibitem{Mizuno09}
S.~Mizuno, F.~Arroja, K.~Koyama, and T.~Tanaka, {\it {Lorentz boost and
  non-Gaussianity in multi-field DBI-inflation}},  {\em Phys. Rev.} {\bf D 80}
  (2009) 023530, [\href{http://xxx.lanl.gov/abs/0905.4557}{{\tt
  arXiv:0905.4557}}].

\bibitem{Langlois09a}
D.~Langlois, S.~Renaux-Petel, and D.~A. Steer, {\it {Multi-field DBI inflation:
  Introducing bulk forms and revisiting the gravitational wave constraints}},
  {\em JCAP} {\bf 0904} (2009) 021,
  [\href{http://xxx.lanl.gov/abs/0902.2941}{{\tt arXiv:0902.2941}}].

\bibitem{Gao09a}
X.~Gao, M.~Li, and C.~Lin, {\it {Primordial Non-Gaussianities from the
  Trispectra in Multiple Field Inflationary Models}},  {\em JCAP} {\bf 0911}
  (2009) 007, [\href{http://xxx.lanl.gov/abs/0906.1345}{{\tt
  arXiv:0906.1345}}].

\bibitem{Cai09}
Y.-F. Cai and W.~Xue, {\it {N-flation from multiple DBI type actions}},  {\em
  Phys. Lett.} {\bf B 680} (2009) 395--398,
  [\href{http://xxx.lanl.gov/abs/0809.4134}{{\tt arXiv:0809.4134}}].

\bibitem{Cai09-1}
Y.-F. Cai and H.-Y. Xia, {\it {Inflation with multiple sound speeds: a model of
  multiple DBI type actions and non-Gaussianities}},  {\em Phys. Lett.} {\bf B
  677} (2009) 226--234, [\href{http://xxx.lanl.gov/abs/0904.0062}{{\tt
  arXiv:0904.0062}}].

\bibitem{Pi11}
S.~Pi and D.~Wang, {\it {Dynamics of Cosmological Perturbations in Multi-Speed
  Inflation}},  \href{http://xxx.lanl.gov/abs/1107.0813}{{\tt
  arXiv:1107.0813}}.

\bibitem{Emery12}
J.~Emery, G.~Tasinato, and D.~Wands, {\it {Local non-Gaussianity from rapidly
  varying sound speeds}},  {\em JCAP} {\bf 1208} (2012) 005,
  [\href{http://xxx.lanl.gov/abs/1203.6625}{{\tt arXiv:1203.6625}}].

\bibitem{Kidani12}
T.~Kidani, K.~Koyama, and S.~Mizuno, {\it {Non-Gaussianities in multi-field DBI
  inflation with a waterfall phase transition}},  {\em Phys.Rev.} {\bf D86}
  (2012) 083503, [\href{http://xxx.lanl.gov/abs/1207.4410}{{\tt
  arXiv:1207.4410}}].

\bibitem{Arnowitt62}
R.~L. Arnowitt, S.~Deser, and C.~W. Misner, {\it {The Dynamics of general
  relativity}},  \href{http://xxx.lanl.gov/abs/gr-qc/0405109}{{\tt
  gr-qc/0405109}}.

\bibitem{Seery05b}
D.~Seery and J.~E. Lidsey, {\it {Primordial non-Gaussianities in single field
  inflation}},  {\em JCAP} {\bf 0506} (2005) 003,
  [\href{http://xxx.lanl.gov/abs/astro-ph/0503692}{{\tt astro-ph/0503692}}].

\bibitem{RenauxPetel09}
S.~Renaux-Petel, {\it {Combined local and equilateral non-Gaussianities from
  multifield DBI inflation}},  {\em JCAP} {\bf 0910} (2009) 012,
  [\href{http://xxx.lanl.gov/abs/0907.2476}{{\tt arXiv:0907.2476}}].

\bibitem{Easson08}
D.~A. Easson, R.~Gregory, D.~F. Mota, G.~Tasinato, and I.~Zavala, {\it
  {Spinflation}},  {\em JCAP} {\bf 0802} (2008) 010,
  [\href{http://xxx.lanl.gov/abs/0709.2666}{{\tt arXiv:0709.2666}}].

\bibitem{Salopek90}
D.~Salopek and J.~Bond, {\it {Nonlinear evolution of long wavelength metric
  fluctuations in inflationary models}},  {\em Phys. Rev.} {\bf D 42} (1990)
  3936--3962.

\bibitem{Kinney97}
W.~H. Kinney, {\it {A Hamilton-Jacobi approach to nonslow roll inflation}},
  {\em Phys. Rev.} {\bf D 56} (1997) 2002--2009,
  [\href{http://xxx.lanl.gov/abs/hep-ph/9702427}{{\tt hep-ph/9702427}}].

\bibitem{Malik01}
K.~A. Malik, {\it {Cosmological perturbations in an inflationary universe}},
  \href{http://xxx.lanl.gov/abs/astro-ph/0101563}{{\tt astro-ph/0101563}}.
  Ph.D. Thesis (Advisor: David Wands).

\bibitem{Malik09}
K.~A. Malik and D.~Wands, {\it {Cosmological perturbations}},  {\em Phys.
  Rept.} {\bf 475} (2009) 1--51, [\href{http://xxx.lanl.gov/abs/0809.4944}{{\tt
  arXiv:0809.4944}}].

\bibitem{Wands10}
D.~Wands, {\it {Local non-Gaussianity from inflation}},  {\em Class. Quant.
  Grav.} {\bf 27} (2010) 124002, [\href{http://xxx.lanl.gov/abs/1004.0818}{{\tt
  arXiv:1004.0818}}].

\bibitem{Starobinskivi85}
A.~A. Starobinsky, {\it {Multicomponent de Sitter (inflationary) stages and the
  generation of perturbations}},  {\em Soviet Journal of Experimental and
  Theoretical Physics Letters} {\bf 42} (1985) 152.

\bibitem{Sasaki96}
M.~Sasaki and E.~D. Stewart, {\it {A General analytic formula for the spectral
  index of the density perturbations produced during inflation}},  {\em Prog.
  Theor. Phys.} {\bf 95} (1996) 71--78,
  [\href{http://xxx.lanl.gov/abs/astro-ph/9507001}{{\tt astro-ph/9507001}}].

\bibitem{Sasaki98}
M.~Sasaki and T.~Tanaka, {\it {Superhorizon scale dynamics of multiscalar
  inflation}},  {\em Prog. Theor. Phys.} {\bf 99} (1998) 763--782,
  [\href{http://xxx.lanl.gov/abs/gr-qc/9801017}{{\tt gr-qc/9801017}}].

\bibitem{Lyth05-1}
D.~H. Lyth and Y.~Rodriguez, {\it {The Inflationary prediction for primordial
  non-Gaussianity}},  {\em Phys. Rev. Lett.} {\bf 95} (2005) 121302,
  [\href{http://xxx.lanl.gov/abs/astro-ph/0504045}{{\tt astro-ph/0504045}}].

\bibitem{Rigopoulos03}
G.~Rigopoulos and E.~Shellard, {\it {The separate universe approach and the
  evolution of nonlinear superhorizon cosmological perturbations}},  {\em Phys.
  Rev.} {\bf D 68} (2003) 123518,
  [\href{http://xxx.lanl.gov/abs/astro-ph/0306620}{{\tt astro-ph/0306620}}].

\bibitem{Byrnes06}
C.~T. Byrnes, M.~Sasaki, and D.~Wands, {\it {The primordial trispectrum from
  inflation}},  {\em Phys. Rev.} {\bf D 74} (2006) 123519,
  [\href{http://xxx.lanl.gov/abs/astro-ph/0611075}{{\tt astro-ph/0611075}}].

\bibitem{Byrnes10a}
C.~T. Byrnes, M.~Gerstenlauer, S.~Nurmi, G.~Tasinato, and D.~Wands, {\it
  {Scale-dependent non-Gaussianity probes inflationary physics}},  {\em JCAP}
  {\bf 1010} (2010) 004, [\href{http://xxx.lanl.gov/abs/1007.4277}{{\tt
  arXiv:1007.4277}}]. * Temporary entry *.

\bibitem{Byrnes10b}
C.~T. Byrnes, S.~Nurmi, G.~Tasinato, and D.~Wands, {\it {Scale dependence of
  local f\_NL}},  {\em JCAP} {\bf 1002} (2010) 034,
  [\href{http://xxx.lanl.gov/abs/0911.2780}{{\tt arXiv:0911.2780}}].

\bibitem{Seery05}
D.~Seery and J.~E. Lidsey, {\it {Primordial non-Gaussianities from
  multiple-field inflation}},  {\em JCAP} {\bf 0509} (2005) 011,
  [\href{http://xxx.lanl.gov/abs/astro-ph/0506056}{{\tt astro-ph/0506056}}].

\bibitem{Chen05a}
X.~Chen, {\it {Running non-Gaussianities in DBI inflation}},  {\em Phys.Rev.}
  {\bf D72} (2005) 123518,
  [\href{http://xxx.lanl.gov/abs/astro-ph/0507053}{{\tt astro-ph/0507053}}].

\bibitem{Komatsu10}
E.~Komatsu {\em et.~al.}, {\it {Seven-Year Wilkinson Microwave Anisotropy Probe
  (WMAP) Observations: Cosmological Interpretation}},  {\em Astrophys. J.
  Suppl.} {\bf 192} (2011) 18, [\href{http://xxx.lanl.gov/abs/1001.4538}{{\tt
  arXiv:1001.4538}}].

\bibitem{Bennett12}
C.~Bennett, D.~Larson, J.~Weiland, N.~Jarosik, G.~Hinshaw, {\em et.~al.}, {\it
  {Nine-Year Wilkinson Microwave Anisotropy Probe (WMAP) Observations: Final
  Maps and Results}},  \href{http://xxx.lanl.gov/abs/1212.5225}{{\tt
  arXiv:1212.5225}}.

\end{thebibliography}
\end{document}